\documentclass[11pt, amsmath, a4paper]{article}

\usepackage{amsfonts}
\usepackage{amssymb}
\usepackage{mathrsfs}
\usepackage{amsmath}
\usepackage{amsbsy}

\newcommand   \Integers {\mathbb Z}
\newcommand   \PositiveIntegers {{\mathbb N}}

\newcommand   \finiteIntegerRing[1] {\basel{\Integers}{{#1}}}
\newcommand   \invertibleRingElements[1] {\basel{\Integers^{\ast}}{{#1}}}
\newcommand  \cnt      {\mathfrak n}
\newcommand  \primeInt  {\mathtt p}
\newcommand  \coprimeInt   {\mathtt q}

\newcommand   \finiteIntegerField {\basel{\mathbb Z}{\primeInt}} 
 
\newcommand   \scalars  {\mathbb F}

\def  \deg  {\small \textsf{deg}}
\def  \mod  {\small{\textsf{ mod }}}

\def  \gcd  {\small \textsf{gcd}}

\newcommand   \powerset {\mathfrak P}

\newcommand  \LookupTable {\textsf{T}}
\newcommand  \SignatureTable {\textsf{S}}
\newcommand  \BackSubstitutionTable {\textsf{B}}
\newcommand  \SignatureVerificationTable {\textsf{V}}
\newcommand  \SignatureAuthenticationTable {\textsf{A}}

\newcommand  \mygrp    {\mathsf G}
\newcommand  \eltSet    {\mathsf E}
\newcommand  \orderofthegroup  {\mathsf n}
\newcommand  \nonzeroscalars {{\mathbb F}^{\ast}}

\newcommand  \false {\mathtt{false}}
\newcommand  \true {\mathtt{true}}

\newcommand  \xx   {\mathbf x}
\newcommand  \yy   {\mathbf y}
\newcommand  \zz   {\mathbf z}

\newcommand  \bolddelta   {\boldsymbol  \delta}
\newcommand  \boldepsilon   {\boldsymbol \epsilon}
\newcommand  \boldomega     {\boldsymbol \omega}
\newcommand  \boldvarepsilon   {\boldsymbol \varepsilon}

\newcommand  \bglb {\big (}
\newcommand  \bgrb {\big )}
\newcommand  \bglc {\big \{}
\newcommand  \bgrc {\big \}}
\newcommand  \bgls {\big [}
\newcommand  \bgrs {\big ]}

\newcommand  \PSPACE {\mathsf {PSPACE}}

\newcommand   \Boolean  {{\scriptstyle{\mathcal B}}\displaystyle{}}
\newcommand   \QBF  {{\scriptstyle{\basel{\Boolean}{\scriptscriptstyle{\mathcal Q}}}}\displaystyle{}}
\newcommand   \QSAT  {{\scriptstyle{\basel{\Boolean}{\scriptscriptstyle{{\mathcal Q}\rm{-SAT}}}}}\displaystyle{}}
\newcommand   \ArithmeticExpressions  {{\scriptstyle{{\mathcal ARITH}\textrm{-}{\mathcal EXP}}}\displaystyle{}}
 
\newcommand   \QuantifiedArithmeticExpressions  {{\scriptstyle{\basel{\ArithmeticExpressions}{\scriptscriptstyle{\mathcal Q}}}}\displaystyle{}}
\newcommand   \SatisfiableQuantifiedArithmeticExpressions  {{\scriptstyle{\basel{\ArithmeticExpressions}{\scriptscriptstyle{\mathcal Q}\scriptscriptstyle{\rm{-SAT}}}}}\displaystyle{}}

\newcommand {\basel}[2]{#1_{_{#2}}}

\newcommand  \polynomials[3] { #1{\mathbf{[}}\basel{#2}{\mathrm{1}},\, \ldots,\, \basel{#2}{\mathit{#3}}{\mathbf{]}} }

\newcommand  \Expressions[3] { {\mathcal {EXP}} {\mathbf{\bglb}} {#1\, ; \, \mathbf{[}}\basel{#2}{\mathrm{1}},\, \ldots,\, \basel{#2}{\mathit{#3}}{\mathbf{]}} {\mathbf{\bgrb}}} 

\newcommand  \multipolynomials[5] { #1{\mathbf{[}}\basel{#2}{\mathrm{1}},\, \ldots,\, \basel{#2}{\mathit{#4}},\, \basel{#3}{\mathrm{1}},\, \ldots,\, \basel{#3}{\mathit{#5}} {\mathbf{]}} }

\newcommand  \multiexpressions[5] {{\mathcal {EXP}} {\mathbf{\bglb}} {#1\,;\, \mathbf{[}}\basel{#2}{\mathrm{1}},\, \ldots,\, \basel{#2}{\mathit{#4}},\, \basel{#3}{\mathrm{1}},\, \ldots,\, \basel{#3}{\mathit{#5}} {\mathbf{]}} {\mathbf{\bgrb}}}

\newcommand  \singlevariablepolynomials[2] {#1{\mathbf{[}#2\mathbf{]}}}

\newcommand  \singlevariableexpressions[2] {{\mathcal {EXP}} {\mathbf{\bglb}} {#1\,;\,{\mathbf{[}#2\mathbf{]}}}{\mathbf{\bgrb}}}

\def \lspace  {\hspace*{-0.1cm}}

\def \tab  {\hspace*{0.5cm}}
\def \ltab  {\hspace*{-0.5cm}}

\newtheorem{theorem}{Theorem}

\newcommand \proof {\rm{\textbf{Proof.}}~~}

\newcommand \qed {\hfill{}$\Box$}

\begin{document}

\title{Multivariate Public Key Cryptography and Digital Signature}

\author{Pulugurtha Krishna Subba Rao\thanks{
Department of Computer Science and Engineering (CSE), Gayatri Vidya Parishad College of Engineering (Autonomous), Madhurawada, VISAKHAPATNAM -- 530 048, Andhra Pradesh, India. E-mail ~: ~ \underline{krishnasubbarao@gvpce.ac.in} } \\
Duggirala Meher Krishna\thanks{Department of Electronics and Communication Engineering (ECE), Gayatri Vidya Parishad College of Engineering (Autonomous), Madhurawada, VISAKHAPATNAM -- 530 048, Andhra Pradesh, India. E-mail~: ~~ \underline{duggiralameherkrishna@gmail.com}} \\
 Duggirala Ravi\thanks{
Department of Computer Science and Engineering (CSE), Gayatri Vidya Parishad College of Engineering (Autonomous), Madhurawada, VISAKHAPATNAM -- 530 048, Andhra Pradesh, India. E-mail ~: ~ \underline{ravi@gvpce.ac.in}} }

\date{}

\maketitle

\begin{abstract}
In this paper, algorithms for multivariate public key cryptography and digital signature are described. Plain messages and encrypted messages are arrays, consisting of elements from a fixed finite ring or field. The encryption and decryption algorithms are based on multivariate mappings. The security of the private key depends on the difficulty of solving a system of parametric simultaneous multivariate equations involving polynomial or exponential mappings. The method is a general purpose utility for most data encryption, digital certificate or digital signature applications. For security protocols of the application layer level in the OSI model, the methods described in this paper are useful.

\paragraph{\rm{\textbf{Mathemetics Subject Classification (2010)}}} {~~03C10,~~11C08, \\
11T71,~~12E20, ~~12Y05,~~13A15,~~13P10,~~81P94,~~94A60
 }
\end{abstract}

\tableofcontents

\section{Introduction}

\subsection{Preliminary Discussion}
The role of cryptographic algorithms is to provide information security  [\cite{Buchmann:2004}, \cite{Koblitz:1994}, \cite{Schneier:1996},  \cite{StallingsCNS:2011},  \cite{StallingsNSE:2011} and  \cite{Stinson:2005}]. In general, proper data encryption and authentication mechanisms with access control are preferred for a trusted secure system [\cite{StallingsCNS:2011} and \cite{StallingsNSE:2011}]. The most popular public key cryptosystems are the RSA [\cite{RSA:1978}], NTRU [\cite{HLS:1999},  \cite{HPS:1998},  \cite{HPS:2001} and  \cite{HS:2001}], ECC  [\cite{Koblitz:1987},  \cite{Miller:1985},  \cite{Smart:1999} and  \cite{Washington:2008}], the algorithms based on diophantine equations [\cite{LCL:1995}] and discrete logarithms [\cite{ElGamal:1985}], and those based on multivariate quadratic polynomials [\cite{BBD:2009} and  \cite{Koblitz:1999}]. The RSA, the NTRU and the ECC are assumed to be secure algorithms unless there are new breakthroughs in integer factoring (for RSA), or in lattice reduction (for NTRU), or in elliptic curve discrete logarithm techniques (for ECC)  [\cite{CS:1997} and \cite{Gentry:2001}].

 In this paper, algorithms for public key cryptography as well as digital signature based on multivariate mappings are described, with plain and encrypted message arrays consisting of elements from a fixed commutative and finite ring or field. The keys can be built up starting from independently chosen small degree polynomial or easy exponential mappings, resulting in fast key generation and facilitating easy changes of keys as often as required. The security depends on the difficulty of solving parametric simultaneous multivariate equations involving polynomial or exponential mappings [\cite{Buchberger:1965},  \cite{CGHMP:2003},  \cite{Faugere:1999},  \cite{Faugere:2002},  \cite{Marker:2000},  \cite{MMP:1996},  \cite{vanDalen:1994} and \cite{vandenDries:2000}] in the case of straightforward attacks, and on the difficulty of finding the private keys in the case of key recovery attacks. For security protocols of the application layer level in the OSI model, the methods described in this paper are useful.

\subsection{\label{Sec-notation}Notation}

 In the sequel, let $\Integers$ be the set of integers, and let
 $\PositiveIntegers$ be the set of positive integers.  For a positive integer $\cnt \geq 2$, let $\finiteIntegerRing{\cnt}$ be the ring of integers with addition and multiplication $\mod \cnt$, and  $\invertibleRingElements{\cnt}$ be the commutative group  of invertible elements in $\finiteIntegerRing{\cnt}$, with respect to multiplication operation in $\finiteIntegerRing{\cnt}$. The representing elements in $\basel{\Integers}{\cnt}$ are taken to be those from the set $\{0,\, \ldots, \cnt-1\} \subseteq \Integers$. Let $\scalars$ be a finite field, consisting of $\primeInt^{n}$ elements for some positive integer $n$ and prime number $\primeInt$, and let $\nonzeroscalars$ be the multiplicative group of nonzero elements in $\scalars$. Let $\mygrp$ be a finite cyclic group of order $\cnt \geq 2$.  Let $\eltSet$ be either $\scalars$ or $\finiteIntegerRing{\cnt}$ or $\mygrp$. If $\eltSet = \mygrp$, where $\mygrp$ is equipped with only the group operation, then $\mygrp$ is isomorphic to $\finiteIntegerRing{\cnt}$, where the group operation in $\mygrp$ is identified with the addition operation of $\finiteIntegerRing{\cnt}$. The addition operation of $\Integers$ is a primary operation, and the multiplication operation, that can be treated as a secondary operation [\cite{Manin:2010}] over the additive group $\Integers$, is defined uniquely by the distribution laws, with $1$ as the multiplicative identity, rendering $\Integers$ as the commutative ring. The same holds for $\finiteIntegerRing{\cnt}$, with $1$ acting as the multiplicative identity. Let $\polynomials{\eltSet}{x}{m}$, for $m \in \PositiveIntegers$, be the algebra of multivariate polynomials in $m$ formal variables $\basel{x}{1},\ldots,\basel{x}{m}$ with coefficients in $\eltSet$.  Now, if $\mygrp = \nonzeroscalars$, for a finite field $\scalars$, then the group operation in $\mygrp$ coincides with the multiplication operation in $\scalars$ and $\polynomials{\mygrp}{x}{m} = \polynomials{\scalars}{x}{m}$. If $m = 1$, then $\polynomials{\eltSet}{x}{m}$ is denoted by $\singlevariablepolynomials{\eltSet}{x}$, with $x = \basel{x}{1}$.  A variable with its name expressed in bold face assumes values from a product space, which is a product of finitely many copies of the same set, and each component of the variable, expressed in the corresponding case without boldness and a positive integer subscript, assumes values from the constituent component space, succinctly as, for example, $\xx = (\basel{x}{1},\, \ldots,\, \basel{x}{m}) \in \eltSet^{m}$,  for some $m \in \PositiveIntegers$.

\subsection{\label{Sec-polynomials-over-Zn}Polynomials over $\finiteIntegerRing{\cnt}$}
  Let $\cnt = \prod_{i = 1}^{r} \basel{\primeInt^{\basel{l}{i}}}{i}$, where $r$ and $\basel{l}{i}$ are positive integers, and $\basel{\primeInt}{i}$ are distinct prime numbers, for $1 \leq i \leq r$. Let $\basel{\coprimeInt}{i} = \basel{\primeInt^{-\basel{l}{i}}}{i}\cnt =
 \prod_{\tiny{\begin{array}{c}
 j = 1\\
 j \neq i
\end{array} } }^{r} \basel{\primeInt^{\basel{l}{j}}}{j} $, 
and let $\basel{m}{i} \in \PositiveIntegers$ be such that 
$\basel{m}{i}\basel{\coprimeInt}{i} \equiv 1 \mod \basel{\primeInt^{\basel{l}{i}}}{i}$, for $1 \leq i \leq r$. Then, $\finiteIntegerRing{\cnt} = \oplus_{i = 1}^{r} \basel{m}{i}\basel{\coprimeInt}{i}\finiteIntegerRing{\basel{\primeInt^{\basel{l}{i}}}{i}}$. 

Now, a polynomial $f(x) \in \singlevariablepolynomials{\basel{\Integers}{\cnt}}{x}$ can be expressed as $\sum_{i = 1}^{r} \basel{m}{i} \basel{\coprimeInt}{i} \basel{f}{i}(x)$, for some unique polynomials $\basel{f}{i}(x) \in \singlevariablepolynomials{\basel{\Integers}{\basel{\primeInt^{\basel{l}{i}}}{i}}}{x}$, for $1 \leq i \leq r$.  For some $x \in \Integers$ and index $i$, where $1 \leq i \leq r$, if $\basel{\primeInt}{i} \mid  f(x)$, then
$\gcd \bglb f(x) \mod \basel{\primeInt^{\basel{l}{i}}}{i}\, , \, \basel{\primeInt}{i}\bgrb$
$=$
$\gcd \bglb \basel{f}{i}(x) \, , \, \basel{\primeInt}{i}\bgrb$
$=$
$\basel{\primeInt}{i} \neq 1$. Thus, $\gcd(f(x), \, \cnt) = 1$, for every $x \in \basel{\Integers}{\cnt}$, if and only if $\gcd(\basel{f}{i}(x),\, \basel{\primeInt}{i}) = 1$, for every $x \in \basel{\Integers}{\basel{\primeInt^{\basel{l}{i}}}{i}}$, for every index $i$, where $1 \leq i \leq r$. Similarly, $f$ is a surjective (hence bijective) mapping from $\basel{\Integers}{\cnt}$ onto $\basel{\Integers}{\cnt}$, if and only if $\basel{f}{i}$ is a surjective (hence bijective) mapping from $\basel{\Integers}{\basel{\primeInt^{\basel{l}{i}}}{i}}$ onto $\basel{\Integers}{\basel{\primeInt^{\basel{l}{i}}}{i}}$,  or equivalently, $\basel{f}{i}(x) \mod \basel{\primeInt}{i}$ is a bijective mapping from $\basel{\Integers}{\basel{\primeInt}{i}}$ into itself and, when $\basel{l}{i} \geq 2$,  $\basel{f'}{i}(x)\not \equiv 0 \mod \basel{\primeInt}{i}$, for all $x \in  \basel{\Integers}{\basel{\primeInt^{\basel{l}{i}}}{i}}$, where $\basel{f'}{i}$ is the formal algebraic derivative of $\basel{f}{i}$, for every index $i$, where $1 \leq i \leq r$ [\cite{LMT:1993}]. Now, if $g(x) \in \singlevariablepolynomials{\basel{\Integers}{\cnt}}{x}$, where $g(x) = \sum_{i = 1}^{r} \basel{m}{i} \basel{\coprimeInt}{i} \basel{g}{i}(x)$, for some $\basel{g}{i}(x) \in \singlevariablepolynomials{\basel{\Integers}{\basel{\primeInt^{\basel{l}{i}}}{i}}}{x}$, for $1 \leq i \leq r$, then $f(x)g(x) = \sum_{i = 1}^{r} \basel{m}{i} \basel{\coprimeInt}{i} \basel{f}{i}(x)\basel{g}{i}(x)$. Thus, (A) $f(x)$ is a unit in $\singlevariablepolynomials{\basel{\Integers}{\cnt}}{x}$, if and only if  $\basel{f}{i}(x)$ is a unit, {\em i.e.},  $\basel{f}{i}(x) \mod \basel{\primeInt}{i} \in  \basel{\Integers^{\ast}}{\basel{\primeInt}{i}}$, for every index $i$, where $1 \leq i \leq r$, (B)  $f(x)$ is reducible in $\singlevariablepolynomials{\basel{\Integers}{\cnt}}{x}$, if and only if $\basel{f}{i}(x)$ is reducible in $\singlevariablepolynomials{\basel{\Integers}{\basel{\primeInt^{\basel{l}{i}}}{i}}}{x}$, for some index $i$, where $1 \leq i \leq r$, and (C) $f(x)$ is irreducible in $\singlevariablepolynomials{\basel{\Integers}{\cnt}}{x}$, if and only if  $\basel{f}{i}(x)$ is irreducible in $\singlevariablepolynomials{\basel{\Integers}{\basel{\primeInt^{\basel{l}{i}}}{i}}}{x}$, or equivalently, $\basel{f}{i}(x) \mod \basel{\primeInt}{i}$ is irreducible in $\singlevariablepolynomials{\basel{\Integers}{\basel{\primeInt}{i}}}{x}$, for every index $i$, where $1 \leq i \leq r$.  Thus, for any positive integer $k$,  $\polynomials{\finiteIntegerRing{\cnt}}{x}{k}$ can be expressed as $\oplus_{i = 1}^{r}\basel{m}{i} \basel{\coprimeInt}{i} \polynomials{\finiteIntegerRing{\basel{\primeInt^{\basel{l}{i}}}{i}}}{x}{k}$.

\subsection{\label{Sec-modular-exponentiation-over-Zn}Modular Exponentiation over $\finiteIntegerRing{\cnt}$}

The modular exponentiation operation is extensively studied in connection with the RSA cryptosystem  [\cite{Buchmann:2004},  \cite{Koblitz:1994},  \cite{RSA:1978},  \cite{Schneier:1996},  \cite{StallingsCNS:2011},  \cite{StallingsNSE:2011} and  \cite{Stinson:2005}]. In this section, the modular exponentiation is extended to the situation, wherein the exponents are functions. The security of the RSA system depends on the difficulty of factorization of a positive integer into its prime factors. However, simplification of computations as well as porting of variables from base level to exponentiation level by a homomorphism requires availability of prime factors in advance for both encryption and decryption, while working with multivariate mappings involving functions as exponents. In the sequel, let  $\varphi$ be Euler phi or totient function [\cite{Buchmann:2004},  \cite{Koblitz:1994},  \cite{Schneier:1996} and  \cite{Stinson:2005}]. Let $\cnt = \prod_{i = 1}^{r} \basel{\primeInt^{\basel{l}{i}}}{i}$, where $r \in \PositiveIntegers$, $\basel{l}{i} \in \PositiveIntegers \backslash \{1\}$ and $\basel{\primeInt}{i}$ are distinct prime numbers, for $1 \leq i \leq r$. Let  $\Expressions{\finiteIntegerRing{\cnt}}{x}{m}$ be the smallest set of  expressions,  closed with respect to addition and multiplication, and containing expressions of the form  $a(\basel{x}{1},\, \ldots,\, \basel{x}{m})^{ b(\basel{x}{1},\, \ldots,\, \basel{x}{m}) }$,  where $a(\basel{x}{1},\, \ldots,\, \basel{x}{m}) $
 $\in$
 $ \polynomials{\finiteIntegerRing{\cnt}}{x}{m}$, and either 
\begin{enumerate}
\item {\label{condition-1-in-modular-exponentiation-over-Zn}}as a formal expression, $b(\basel{x}{1},\, \ldots,\, \basel{x}{m})$ does not depend on $(\basel{x}{1},\, \ldots,\, \basel{x}{m})$ and evaluates to any fixed positive integer, or

\item   $a(\basel{x}{1},\, \ldots,\, \basel{x}{m})$ evaluates to elements in $\invertibleRingElements{\cnt}$, for all values of $(\basel{x}{1},\, \ldots,\, \basel{x}{m})$ in some domain of interest, which is a subset of $\finiteIntegerRing{\cnt}^{m}$,  and  $b(\basel{x}{1},\, \ldots,\, \basel{x}{m})$ is of the form $c(h(\basel{x}{1}),\, \ldots,\, h(\basel{x}{m}))$, for some expression  $c(\basel{z}{1},\, \ldots,\, \basel{z}{m})$
$ \in $
$\Expressions{\finiteIntegerRing{\varphi(\cnt)}}{z}{m}$ and ring homomorphism $h$ from $\finiteIntegerRing{\cnt}$ into $\finiteIntegerRing{\varphi(\cnt)}$.  

\end{enumerate}
\noindent The condition in (\ref{condition-1-in-modular-exponentiation-over-Zn}) above implies that  {\small{$\polynomials{\finiteIntegerRing{\cnt}}{x}{m} \subseteq  \Expressions{\finiteIntegerRing{\cnt}}{x}{m}$}}. Thus, the integers in $\Integers$ and those in $\finiteIntegerRing{\cnt}$, for various modulus positive integers $\cnt \geq 2$, need to be distinguished clearly as separate elements. The expressions in $\Expressions{\finiteIntegerRing{\cnt}}{x}{m}$ are turned into mappings, by identifying appropriate domains of values and interpretation for variables and operations in the respective domains [\cite{vanDalen:1994},  \cite{vandenDries:2000},  \cite{Manin:2010} and  \cite{Marker:2000}]. For $\xx \in \finiteIntegerRing{\cnt}^{m}$ and $s \in \PositiveIntegers \backslash \{1\}$, such that $s \mid \cnt$, let $\xx \mod s = \bglb \basel{x}{1} \mod s,\, \ldots,\, \basel{x}{m} \mod s \bgrb$.  Let  $f(\xx) \in \polynomials{\basel{\Integers}{\cnt}}{x}{m}$ be  such that $f(\xx)$ evaluates to elements in $\invertibleRingElements{\cnt}$, for $\xx \in  X$, for some  $X \subseteq \finiteIntegerRing{\cnt}^{m}$, and let $\basel{f}{i}(\xx) \in \polynomials{\basel{\Integers}{\basel{\primeInt^{\basel{l}{i}}}{i}}}{x}{m}$, for $1 \leq i \leq r$, be such that $f(\xx) = \sum_{i = 1}^{r} \basel{m}{i} \basel{\coprimeInt}{i} \basel{f}{i}(\xx  \mod \basel{\primeInt^{\basel{l}{i}}}{i})$. Now, for $\xx \in X$ and  $k \in \Integers$, the following holds: $\bglb f(\xx)\bgrb^{k}$
 $ =$
$\bglb f(\xx)\bgrb^{k \mod \varphi(\cnt)}$
 $=$
$  \sum_{i = 1}^{r} \basel{m}{i} \basel{\coprimeInt}{i} \bglb\basel{f}{i}(\xx \mod \basel{\primeInt^{\basel{l}{i}}}{i})\bgrb^{k \mod \varphi(\cnt)}$
 $=$
$  \sum_{i = 1}^{r} \basel{m}{i} \basel{\coprimeInt}{i} \bglb\basel{f}{i}(\xx \mod \basel{\primeInt^{\basel{l}{i}}}{i})\bgrb^{k \mod \varphi(\basel{\primeInt^{\basel{l}{i}}}{i})}$. Let $g(\yy) \in \polynomials{\basel{\Integers}{\varphi(\basel{\Integers}{\cnt})}}{y}{n}$ and $\basel{g}{i}(\zz) \in \polynomials{\basel{\Integers}{\varphi(\basel{\primeInt^{\basel{l}{i}}}{i})}}{z}{n}$ be such that the following holds:
 $ \basel{g}{i}\bglb\yy \mod \varphi(\basel{\primeInt^{\basel{l}{i}}}{i})\bgrb = g(\yy) \mod \varphi\bglb \basel{\primeInt^{\basel{l}{i}}}{i} \bgrb $, for $1 \leq i \leq r$. 
Thus, $f^{g(\yy)}(\xx) = \sum_{i = 1}^{r}  \basel{m}{i} \basel{\coprimeInt}{i} \basel{f^{g(\yy )}}{i}(\xx) = \sum_{i = 1}^{r}  \basel{m}{i} \basel{\coprimeInt}{i} \basel{f^{\basel{g}{i}(\yy \mod \varphi(\basel{\primeInt^{\basel{l}{i}}}{i}))}}{i}(\xx\mod \basel{\primeInt^{\basel{l}{i}}}{i})$, for independent vectors $\xx \in X$ and $\yy \in \finiteIntegerRing{\varphi(\cnt)}^{n}$. Now, $\varphi(\basel{\primeInt^{\basel{l}{i}}}{i}) = (\basel{\primeInt}{i}-1)\basel{\primeInt^{\basel{l}{i}-1}}{i}$, where $\basel{l}{i} \geq 2$,  for $1 \leq i \leq r$. Let $\basel{w}{i} = (\basel{\primeInt}{i}-1)^{-1} \mod \basel{\primeInt^{\basel{l}{i}-1}}{i}$, and let $\basel{h}{i}\,:\, \finiteIntegerRing{\basel{\primeInt^{\basel{l}{i}}}{i}} \rightarrow  \finiteIntegerRing{\varphi(\basel{\primeInt^{\basel{l}{i}}}{i})}$ be the map defined by $\basel{h}{i}(x) = (\basel{\primeInt}{i}-1)(\basel{w}{i} x \mod \basel{\primeInt^{\basel{l}{i}-1}}{i})$, for $1 \leq i \leq r$. Then, $\basel{h}{i}$  is a ring homomorphism, for $1 \leq i \leq r$. Now,  let $h\bglb\sum_{i =1}^{r}\basel{m}{i}\basel{\coprimeInt}{i}\basel{z}{i}\bgrb = \bglb  \basel{h}{1}(\basel{z}{1}),\, \ldots, \,  \basel{h}{r}(\basel{z}{r})\bgrb$, for $\basel{z}{i} \in \finiteIntegerRing{\basel{\primeInt^{\basel{l}{i}}}{i}}$ and $1 \leq i \leq r$. Then, the map $h$ is a ring homomorphism from the ring $\oplus_{i = 1}^{r} \basel{m}{i} \basel{\coprimeInt}{i}\finiteIntegerRing{\basel{\primeInt^{\basel{l}{i}}}{i}}$ into the ring of direct product $\prod_{i = 1}^{r} \finiteIntegerRing{\varphi(\basel{\primeInt^{\basel{l}{i}}}{i})}$. If the base level and exponentiation level interpretation maps are $\basel{\mathcal I}{\textrm{base}}$ and  $\basel{\mathcal I}{\mathrm{exponent}}$, respectively, then $\basel{\mathcal I}{\mathrm{exponent}}$ can be chosen to be $h \circ \basel{\mathcal I}{\mathrm{base}}$, applied from right to left in the written order, preserving the respective ring operations in the base level and exponentiation level subexpressions. If $\basel{l}{i} = 1$, for some index $i$, where $1\leq i \leq r$, then exponentiation along $i$th component can be carried by interpreting $\finiteIntegerRing{\basel{\primeInt}{i}}$ to be a finite field, and porting values of base level  expressions to exponentiation level expressions by discrete logarithm mapping, as discussed in section \ref{Sec-modular-exponentiation-over-Finite-Fields}.

\subsection{\label{Sec-modular-exponentiation-over-Finite-Fields}Modular Exponentiation over $\scalars$}

  Let $\scalars$ be a finite field containing $\primeInt^{n}$ elements and $\cnt = \primeInt^{n}-1$, for some prime number $\primeInt$ and positive integer $n$. Let  $\Expressions{\scalars}{x}{m}$ be the smallest set of  expressions,  closed with respect to addition and multiplication, and containing expressions of the form  $a(\basel{x}{1},\, \ldots,\, \basel{x}{m})^{ b(\basel{x}{1},\, \ldots,\, \basel{x}{m}) }$,  where $a(\basel{x}{1},\, \ldots,\, \basel{x}{m}) $
 $\in$
 $ \polynomials{\scalars}{x}{m}$, and either  
\begin{enumerate}
\item \label{item-1-in-modular-exponentiation-over-Finite-Fields}as a formal expression, $b(\basel{x}{1},\, \ldots,\, \basel{x}{m})$ does not depend on $(\basel{x}{1},\, \ldots,\, \basel{x}{m})$ and evaluates to any fixed positive integer, or 
\item  \label{item-2-in-modular-exponentiation-over-Finite-Fields} $a(\basel{x}{1},\, \ldots,\, \basel{x}{m})$ evaluates to elements in $\nonzeroscalars$, for all values of $(\basel{x}{1},\, \ldots,\, \basel{x}{m})$ in some domain of interest, which is a subset of $\mygrp^{m}$, where $\mygrp = \nonzeroscalars$, and  $b(\basel{x}{1},\, \ldots,\, \basel{x}{m})$ is of the form $c(h(\basel{x}{1}),\, \ldots,\, h(\basel{x}{m}))$, for some expression 
{\small{ $c(\basel{z}{1},\, \ldots,\, \basel{z}{m})$
 $ \in $
 $\Expressions{\finiteIntegerRing{\cnt}}{z}{m}$ and group isomorphism $h$ from $\mygrp$ into $\finiteIntegerRing{\cnt}$}.}    
\end{enumerate}
The condition in (\ref{item-1-in-modular-exponentiation-over-Finite-Fields}) above implies that {\small{$\polynomials{\scalars}{x}{m} \subseteq \Expressions{\scalars}{x}{m}$}}. For a primitive element $a \in \nonzeroscalars$, let $\basel{\log}{a} : \nonzeroscalars \rightarrow \finiteIntegerRing{\cnt}$ be the discrete logarithm function defined by $\basel{\log}{a}(g) = x$, exactly when $a^{x} = g$, for $g \in \nonzeroscalars$ and $x \in \finiteIntegerRing{\cnt}$. Thus, the group homomorphism $h$ can be taken to be $\basel{\log}{a}$. If the base level and exponentiation level interpretation maps are $\basel{\mathcal I}{\textrm{base}}$ and  $\basel{\mathcal I}{\textrm{exponent}}$, respectively, then  $\basel{\mathcal I}{\textrm{exponent}}$ can be chosen to be $\basel{\log}{a} \circ ~ \basel{\mathcal I}{\textrm{base}}$, applied from right to left in the written order. For porting a subexpression involving addition operation in $\scalars$, such as, for example,  $f(\xx) \in \polynomials{\scalars}{x}{m}$, where $f(\xx) \neq 0$, for $\xx \in \mygrp^{m}$, where $\mygrp = \nonzeroscalars$, occurring in a base level expression to an exponentiation level,  the base level subexpression is replaced by a supplementary variable $z$, which is ported to first exponentiation level by the discrete logarithm mapping. In the subsequent levels of exponentiation, the interpretation is performed by applying ring homomorphisms, as  discussed in section \ref{Sec-modular-exponentiation-over-Zn}.

\section{Main Results}

\subsection{\label{Sec-parametric-injective-mappings}Parametric Injective Mappings}

 Let $\eltSet$ be either $\scalars$ or $\finiteIntegerRing{\cnt}$.  Let $\mygrp \subseteq \eltSet$ be the domain of interpretation for the variables occurring in the mappings.   For  $l \in \{0\} \cup \PositiveIntegers$ and  $m \in \PositiveIntegers$, a parametric multivariate injective mapping $\eta\bglb \basel{z}{1}, \, \ldots,\, \basel{z}{l};\, (\basel{x}{1},\,\ldots,\, \basel{x}{m})\bgrb$ from $\mygrp^{m}$ into $\eltSet^{m}$ is a multivariate injective mapping, which is an expression from either {\small{$\multipolynomials{\eltSet}{x}{z}{m}{l}$}} or {\small{$\multiexpressions{\eltSet}{x}{z}{m}{l}$}} with interpretation conventions as discussed in sections \ref{Sec-modular-exponentiation-over-Zn} and \ref{Sec-modular-exponentiation-over-Finite-Fields}, as appropriate, for $(\basel{x}{1},\, \ldots,\, \basel{x}{m}) \in \mygrp^{m}$ and  $(\basel{z}{1}, \, \ldots,\, \basel{z}{l}) \in {\mathcal Z} \subseteq \eltSet^{l}$,  and its parametric inverse $\eta^{-1}\bglb \basel{z}{1}, \, \ldots,\, \basel{z}{l};\, (\basel{y}{1},\,\ldots,\, \basel{y}{m})\bgrb$ is such that, for every fixed $(\basel{z}{1}, \, \ldots,\, \basel{z}{l}) \in {\mathcal Z} $, the following holds:  if $\eta\bglb\basel{z}{1}, \, \ldots,\, \basel{z}{l};\, (\basel{x}{1},\,\ldots,\, \basel{x}{m})\bgrb$
 $ = $
 $(\basel{y}{1},\, \ldots,\, \basel{y}{m})$, then $(\basel{x}{1},\,\ldots,\, \basel{x}{m})$
 $ = $
 $ \eta^{-1}\bglb\basel{z}{1}, \, \ldots,\, \basel{z}{l};\, $
 $(\basel{y}{1},\, \ldots,\, \basel{y}{m})\bgrb$, for every $(\basel{x}{1},\, \ldots,\, \basel{x}{m})$
 $ \in $
 $\mygrp^{m}$ and  $(\basel{y}{1},\, \ldots,\, \basel{y}{m}) $
 $ \in $
 $\eltSet^{m}$. For example, let $\cnt$ be the set cardinality of $\mygrp = \nonzeroscalars$, $a \in \nonzeroscalars$ be a fixed primitive element, which is made known in the public key, and $\eta\bglb\basel{z}{1}, \, \ldots,\, \basel{z}{l};\, x \bgrb$
 $ = $
 $f(\basel{z}{1},\, \ldots,\, \basel{z}{l})x^{g(\basel{\log}{a}(\basel{z}{1}),\, \ldots,\, \basel{\log}{a}(\basel{z}{l}))}$, where {\small{$f(\basel{z}{1},\, \ldots,\, \basel{z}{l}) \in \Expressions{\scalars}{z}{l}$}} and {\small{$g(\basel{t}{1},\, \ldots,\, \basel{t}{l})\in \Expressions{\basel{\Integers}{\orderofthegroup}}{t}{l}$}} are such that {\small{$f(\basel{z}{1},\, \ldots,\, \basel{z}{l}) \neq 0$}}, for {\small{$\basel{z}{1},\, \ldots,\, \basel{z}{l}$
 $ \in $
 $\nonzeroscalars$}}, and {\small{$\gcd\bglb g(\basel{t}{1},\, \ldots,\, \basel{t}{l}),\, \cnt) $
 $ = $
 $ 1$}}, for {\small{$\basel{t}{1},\, \ldots,\, \basel{t}{l} $
 $\in $
 $\basel{\Integers}{\cnt}$}}. Then, {\small{$\eta\bglb\basel{z}{1}, \, \ldots,\, \basel{z}{l};\, x\bgrb$}} is a parametric bijective mapping from $\nonzeroscalars$ into $\nonzeroscalars$, with $\basel{z}{1},\, \ldots,\, \basel{z}{l} \in \nonzeroscalars$ as parameters, and the inverse mapping of $\eta$ is 
 {\small{$\eta^{-1}\bglb\basel{z}{1}, \, \ldots,\, \basel{z}{l};\, x\bgrb $
 $ = $
 $[~[f(\basel{z}{1},\, \ldots,\, \basel{z}{l})]^{-1}x~]^{[~[g(\basel{\log}{a}(\basel{z}{1}),\, \ldots,\, \basel{\log}{a}(\basel{z}{l}))]^{-1}\mod \cnt~]}$}}.

 For the multivariate surjective mappings for digital signature scheme 
 discussed at the end of section \ref{Sec-PKC-and-DS}, mappings 
{\small{$f(\basel{z}{1},\, \ldots,\, \basel{z}{l})$ 
$\in $
$\Expressions{\scalars}{z}{l}$}} and
{\small{$g(\basel{t}{1},\, \ldots,\, \basel{t}{l})$
$\in $
$\Expressions{\basel{\Integers}{\orderofthegroup}}{t}{l}$}}
can be chosen, such that both the conditions 
{\small{$f(\basel{z}{1},\, \ldots,\, \basel{z}{l})$
$ \neq $
$0$}} and
{\small{$\gcd\bglb g(\basel{\log}{a}(\basel{z}{1}),\, \ldots,\, \basel{\log}{a}(\basel{z}{l})),\, \cnt) $
 $ = $
 $ 1$}}, simultaneously hold for {\small{$(\basel{z}{1},\, \ldots,\, \basel{z}{l})$
$\in$
${\mathcal Z} \subseteq \mygrp^{l}$},}
where the required exact domain ${\mathcal Z} \neq \emptyset$
is a private key and known only to the signer.
 
\subsubsection{\label{sec-parametrization-of-permutation-polynomials}Parametrization Methods}

Let, for some positive integers $k$, $l$ and $m$, 
$\basel{g}{i}\bglb\basel{z}{1},\, \ldots,\, \basel{z}{l}\bgrb$, $1 \leq i \leq k$,
be a partition of unity of $\eltSet^{l}$, {\em i.e.},
$\sum_{i = 1}^{k} \basel{g}{i}\bglb\basel{z}{1},\, \ldots,\, \basel{z}{l}\bgrb = 1$
and $ \basel{g}{i}\bglb\basel{z}{1},\, \ldots,\, \basel{z}{l}\bgrb
\cdot \basel{g}{j}\bglb\basel{z}{1},\, \ldots,\, \basel{z}{l}\bgrb$
$  = 0$,
$i \neq j$, $1 \leq i,\, j \leq k$, for every
$\bglb\basel{z}{1},\, \ldots,\, \basel{z}{l}\bgrb \in \eltSet^{l}$. The partition of unity required for the parametric mappings discussed of this section need not necessarily be strict, and it is possible that, for some $i$, where $1 \leq i \leq k$, $ \basel{g}{i}\bglb\basel{z}{1},\, \ldots,\, \basel{z}{l}\bgrb = 0$, for every  $\bglb\basel{z}{1},\, \ldots,\, \basel{z}{l}\bgrb \in \eltSet^{l}$.
Let $\basel{\zeta}{i}\bglb \basel{z}{1},\, \ldots,\, \basel{z}{l};\, \xx \bgrb$,
$1 \leq i \leq k$, $\xx = (\basel{x}{1}, \ldots,\, \basel{x}{m})$,
be parametric multivariate injective mappings from $\mygrp^{m}$ into $\eltSet^{m}$, that may or may not depend on the parameters $ \basel{z}{1},\, \ldots,\, \basel{z}{l}$. The vectors $\xx$ and $\basel{\zeta}{i}\bglb \basel{z}{1},\, \ldots,\, \basel{z}{l};\, \xx \bgrb$, $1 \leq i \leq k$, are identified with the corresponding $m \times 1$ column vectors, whose $j$-th row entry is the $j$-th element, for $1 \leq j \leq m$, for allowing them to become amenable to matrix operations. Let
 $\basel{\phi}{i}\bglb \basel{z}{1},\, \ldots,\, \basel{z}{l}\bgrb$ be an $m \times m$ matrix, and
$\, \basel{\chi}{i}\bglb \basel{z}{1},\, \ldots,\, \basel{z}{l}\bgrb $ 
 be $m \times 1$ vectors, both with multivariate expressions as entries, such that 
$\basel{\phi}{i}\bglb \basel{z}{1},\, \ldots,\, \basel{z}{l}\bgrb$ evaluates to an invertible matrix, for every $\bglb\basel{z}{1},\, \ldots,\, \basel{z}{l}\bgrb \in \eltSet^{l}$ and $1 \leq i \leq k$. Then, the expression
{\small{$\eta (\basel{z}{1},\, \ldots,\, \basel{z}{l};\, \xx)$}}
$ =$
{\small{$\sum_{i = 1}^{k} \basel{g}{i}(\basel{z}{1},\, \ldots,\, \basel{z}{l}) \cdot
\basel{\phi}{i}(\basel{z}{1},\, \ldots,\, \basel{z}{l}) \cdot
[\basel{\zeta}{i}(\basel{z}{1},\, \ldots,\, \basel{z}{l};\, \xx) +
\basel{\chi}{i}(\basel{z}{1},\, \ldots,\, \basel{z}{l})]$}}
is a parametric multivariate injective mapping, with its
parametric inverse
{\small{$\eta^{-1}(\basel{z}{1},\, \ldots,\, \basel{z}{l};\, \yy)$}}
$ =$
{\small{$\sum_{i = 1}^{k} \basel{g}{i}(\basel{z}{1},\, \ldots,\, \basel{z}{l}) \cdot
\basel{\zeta^{-1}}{i}(\basel{z}{1},\, \ldots,\, \basel{z}{l};\,  \basel{\xx}{i})$}\,,}
where 
{\small{ $\basel{\xx}{i} = \bgls [\basel{\phi}{i}(\basel{z}{1},\, \ldots,\, \basel{z}{l})]^{-1}
 \cdot \yy \bgrs - \basel{\chi}{i}(\basel{z}{1},\, \ldots,\, \basel{z}{l}) $},} 
 {\small{$\yy = (\basel{y}{1},\, \ldots, \,  \basel{y}{m})$},} which is also identified with the corresponding
 $m \times 1$ column vector, and {\small{$ [\basel{\phi}{i}(\basel{z}{1},\, \ldots,\, \basel{z}{l})]^{-1}$}} is the matrix inverse of {\small{$ [\basel{\phi}{i}(\basel{z}{1},\, \ldots,\, \basel{z}{l})]$},} for $1 \leq i \leq k$.

 For the multivariate surjective mappings for digital signature scheme 
 discussed at the end of section \ref{Sec-PKC-and-DS}, it is
 possible to choose 
  $\basel{\zeta}{i}\bglb \basel{z}{1},\, \ldots,\, \basel{z}{l};\, \xx \bgrb$
 to be bijective, only for some indexes $i$, where $1 \leq i \leq k$, 
 letting it be arbitrary for the remaining indexes. Since the domain information
 is a private key, as discussed in the last paragraph of the preceding section,
 the updates mentioned here must be so chosen that the effective domain will become
 feasible, while maintaining it as a private key.

\subsubsection{\label{Sec-partition-of-unity}Partition of Unity of $\scalars$}

 Let $f(z) \in \singlevariableexpressions{\scalars}{z}$, which is
called a discriminating function, and let $\basel{K}{f}$ be the
codomain of $f$, {\em i.e.}, $\basel{K}{f} =
 \{f(x)\,:\, x \in \scalars\} = \{\basel{a}{i}\,:\, 1 \leq i \leq k\}$,
 for some positive integer $k$.
Let {\small{$\basel{\ell}{i}(x) = \bigg[\prod_{\tiny{
 \begin{array}{c}   j = 1\\  j \neq i \end{array} } }^{k} 
\bglb \basel{a}{i}-\basel{a}{j}\bgrb\bigg]^{-1}\cdot
  \prod_{\tiny{
 \begin{array}{c}  j = 1\\  j \neq i \end{array} } }^{k} 
\bglb f(x)-\basel{a}{j}\bgrb$},} $1 \leq i \leq k$.
Then, $\basel{\ell}{i}(x) = 1$, for $x \in \basel{E}{i} = \{z \in \scalars\,:\, f(z) - \basel{a}{i} = 0\}$,
and $\basel{\ell}{i}(x) = 0$, for $x \in \scalars \backslash \basel{E}{i}$, $1 \leq i \leq k$.
Thus, $\{\basel{E}{i}\,:\,1 \leq i \leq k\}$ is a partition of $\scalars$, and $\basel{\ell}{i}(x)$
is the characteristic function of the equivalence class $\basel{E}{i}$, $1 \leq i \leq k$.
Now, the set
 $\{ \basel{g}{i}(\basel{z}{1},\, \ldots,\, \basel{z}{l})
= \basel{\ell}{i}\bglb h(\basel{z}{1},\, \ldots,\, \basel{z}{l})\bgrb\,:\,
 1 \leq i \leq k\}$, where
 $h(\basel{z}{1},\, \ldots,\, \basel{z}{l}) \in \Expressions{\scalars}{z}{{\mathit l}}$,
is a partition of unity of $\scalars^{l}$.
\\

\begin{small}
\noindent{\bf{Examples.}}~   (A) ~  Let the vector space dimension of $\scalars$ be $n$
 as an extension field of $\basel{\Integers}{\primeInt}$, and
 let $f(z) = \sum_{i = 1}^{n}\basel{a}{i}z^{\primeInt^{i-1}}$, 
 where $\basel{a}{i} \in \scalars$, $1 \leq i \leq n$,
 be a noninvertible linear operator from $\scalars$ into $\scalars$,
 with $\basel{\Integers}{\primeInt}$ as the field.
 For every linear operator $T$ from $\scalars$ into $\scalars$
 with $\basel{\Integers}{\primeInt}$ as the field, there exist
 scalars $\basel{c}{i} \in \scalars$, $1 \leq i \leq n$,
 such that $Tz = \sum_{i=1}^{n}\basel{c}{i}z^{\primeInt^{i-1}}$
 [\cite{LN:1986}]. Now, each equivalence class is an affine
 vector subspace of the form $\{y+x\,:\, f(x) = 0, ~x \in \scalars\}$,
 for some $y \in \scalars$. Thus, if $r$ is the rank of $f$
 as linear operator from $\scalars$ into $\scalars$ with
 $\basel{\Integers}{\primeInt}$ as the field, then the nullity
 of $f$ is $n-r$, each equivalence class has $\primeInt^{n-r}$
 elements, and there are $k = \primeInt^{r}$ equivalence classes.
 For the number of equivalence classes to be small, the rank $r$
 of $f$ must be small, such as $r = 1$ or $r = 2$.
 ~~(B) ~ Let $f(z) = z^{r}$,  where  $r$ is a large positive integer
 dividing $\primeInt^{n}-1$.  Now, the equivalence classes are $\{0\}$
 and the cosets of the congruence relation $x \sim y$ if and only if
 $( x^{-1}y )^{r} = 1$, for $x,\, y \in \scalars \backslash \{0\}$.
 Since $\basel{K}{f} = \{0\} \cup \{z^{r}\,:\, z \in \scalars \backslash \{0\}\}$, 
 there are $k = 1+(\primeInt^{n}-1)/r$ equivalence classes. 
 
\end{small}

\subsubsection{\label{partition-of-unity-of-Zn}Partition of Unity of $\basel{\Integers}{\primeInt^{l}}$}

Let $s \in \PositiveIntegers$ be a divisor of $(\primeInt-1)$  and $k = 1+\frac{(\primeInt-1)}{s}$. Now, $\primeInt^{l-1} \geq l$, for any $l \in \PositiveIntegers$ and prime number $\primeInt$. Let $h(x) = x^{s\primeInt^{l-1}}$, for $x \in \basel{\Integers}{\primeInt^{l}}$. Then, $\bglb h(x) \bgrb^{k-1} = 1$, for $x \in \basel{\Integers^{\star}}{\primeInt^{l}}$, and $h(x) = 0$, for $x \in \basel{\Integers}{\primeInt^{l}} \backslash \basel{\Integers^{\star}}{\primeInt^{l}}$.  Thus, the set $\{x^{s\primeInt^{l-1}}\,:\, x \in \basel{\Integers}{\primeInt^{l}}\}$ contains $k$ distinct elements. Let $x, \, y \in \basel{\Integers}{\primeInt^{l}}$ be such that $h(x) \neq h(y)$. If $h(x) = 0$ or $h(y) = 0$, then $(h(y)-h(x))  \in \basel{\Integers^{\star}}{\primeInt^{l}}$. Now, let $x,\, y \in \basel{\Integers^{\star}}{\primeInt^{l}}$. If $(x^{-1}y)^{s\primeInt^{l-1}} = 1+b\primeInt^{t}$, for some $b \in \basel{\Integers^{\star}}{\primeInt^{l}}$ and $t \in \PositiveIntegers$, then, since  $1 + b \primeInt^{t} \sum_{i = 1}^{k-1} \frac{(k-1)!}{i!(k-i-1)!}  b^{i-1}\primeInt^{(i-1)t} = 
(1+b\primeInt^{t})^{k-1} = \bglb(x^{-1}y)^{s\primeInt^{l-1}}\bgrb^{k-1} = 1 \mod \primeInt^{l}$, it follows that either $t \geq l$ or  $(k-1)+\sum_{i = 2}^{k-1} \frac{(k-1)!}{i!(k-i-1)!} b^{i-1}\primeInt^{(i-1)t} = 0 \mod \primeInt^{l-t}$. However, since $k = 1+\frac{\primeInt-1}{s}$, and therefore, $1 \leq k-1 \leq \primeInt-1$, it follows that $(k-1)+\sum_{i = 2}^{k-1} \frac{(k-1)!}{i!(k-i-1)!} b^{i-1}\primeInt^{(i-1)t} = k-1 \mod \primeInt$. Thus, if  $x,\, y \in \basel{\Integers^{\star}}{\primeInt^{l}}$ and 
$h(x) \neq h(y)$, then $(x^{-1}y)^{s\primeInt^{l-1}} - 1  \neq 0 \mod\primeInt$, and hence if  $x,\, y \in \basel{\Integers}{\primeInt^{l}}$ and  $h(x) \neq h(y)$, then $(h(y)-h(x)) \in \basel{\Integers^{\star}}{\primeInt^{l}}$. If $\basel{a}{j} \in \basel{\Integers}{\primeInt^{l}}$, $1 \leq j \leq k$, are such that $\{x^{s\primeInt^{l-1}}\,:\, x \in \basel{\Integers}{\primeInt^{l}}\} = \{\basel{a}{j}\,:\, 1 \leq j \leq k\}$, then $(\basel{a}{i}-\basel{a}{j}) \in \basel{\Integers^{\star}}{\primeInt^{l}}$, for $i \neq j$, $1 \leq i,\, j \leq k$, and the Lagrange interpolation polynomials  $\basel{g}{j}(x) \in \singlevariablepolynomials{\basel{\Integers}{\primeInt}}{x}$ can be obtained for the equivalence classes $\basel{E}{j} = \{x^{s\primeInt^{l-1}} = \basel{a}{j}\,:\, x \in \basel{\Integers}{\primeInt^{l}}\}$.  Thus, corresponding to every homomorphism of $\basel{\Integers^{\star}}{\primeInt}$ into $\basel{\Integers^{\star}}{\primeInt}$, a partition of unity of $\basel{\Integers}{\primeInt^{l}}$ can be obtained.

\subsubsection{\label{Sec-multivariate-polynomials-that-evaluate-to-only-invertible-elements}Multivariate Mappings that Evaluate to only Invertible Elements}

Let $f(z) \in \singlevariablepolynomials{\scalars}{z}$ be a polynomial which is not surjective as a mapping from $\scalars$ into $\scalars$. Then, there exists an element $c \in \scalars$, such that $f(z)-c \neq 0$, for every $z \in \scalars$. For $a \in \scalars \backslash \{0\}$ and $g(\basel{z}{1},\, \ldots,\, \basel{z}{l}) \in \polynomials{\scalars}{z}{l}$,  $a\bglb f({\small{g(\basel{z}{1},\, \ldots,\, \basel{z}{l})}}) -c \bgrb \neq 0$, for every $(\basel{z}{1},\, \ldots,\, \basel{z}{l}) \in \scalars^{l}$.
\\

\begin{small}
\noindent{\bf{Examples.}}~ (A)~ Let $f(z)$ be a product of irreducible polynomials
in $\singlevariablepolynomials{\scalars}{z}$ of degree $2$ or more each. 
Then, $c$ can be chosen to be $0$.~~
(B)~ Let the vector space dimension of $\scalars$ be $n$
as an extension field of $\basel{\Integers}{\primeInt}$, and
let $f(z) = \sum_{i = 1}^{n}\basel{a}{i}z^{\primeInt^{i-1}}$, 
where $\basel{a}{i} \in \scalars$, $1 \leq i \leq n$,
be a noninvertible linear operator from $\scalars$ into $\scalars$,
with $\basel{\Integers}{\primeInt}$ as the field. Then,
for any basis $\{\basel{\alpha}{1},\, \ldots,\, \basel{\alpha}{n} \}$
for $\scalars$, with $\basel{\Integers}{\primeInt}$ as the field,
there exists an index $j$, $1 \leq j \leq n$, such that 
$\sum_{i = 1}^{n}\basel{a}{i}z^{\primeInt^{i-1}} - \basel{\alpha}{j}  
\neq 0$, for every $z \in \scalars$, and $c$ can be taken
to be $\basel{\alpha}{j}$.~~
(C) ~ Let $r \geq 2$ be a positive integer divisor of $\primeInt^{n}-1$,
and let $f(z) = z^{r}$. Then, there exists an element
$c \in \scalars \backslash \{0\}$, such that $c^{(\primeInt^{n}-1)/r} \neq 1$.
Now, since $c^{(\primeInt^{n}-1)/r} \neq 0$ and
$c^{(\primeInt^{n}-1)/r} \neq 1$, it follows that
 $f(z) - c \neq 0$, for every $z \in \scalars$.
\\

\end{small}

If $f(z) \in \singlevariablepolynomials{\scalars}{z}$
is such that $f(z) \neq 0$, for every $z \in \scalars$, then
$[f(z)]^{-1} = \sum_{i = 1}^{k} \basel{a^{-1}}{i}\basel{\ell}{i}(z)$, where $\{\basel{a}{i}\,:\, 1 \leq i \leq k\}  = 
 \{f(z)\,:\, z \in \scalars\} $, and $\basel{\ell}{i}(z) = 
\bigg[\prod_{\tiny{ \begin{array}{c}   j = 1\\  j \neq i \end{array} } }^{k} 
\bglb \basel{a}{i}-\basel{a}{j}\bgrb\bigg]^{-1}\cdot
  \prod_{\tiny{
 \begin{array}{c}  j = 1\\  j \neq i \end{array} } }^{k} 
\bglb f(z)-\basel{a}{j}\bgrb$, $1 \leq i \leq k$.

Let $\cnt = \prod_{i = 1}^{r} \basel{\primeInt^{\basel{l}{i}}}{i}$, where $r,\, \basel{l}{i} \in \PositiveIntegers$ and $\basel{\primeInt}{i}$ are distinct prime numbers, for $1 \leq i \leq r$,  and  $f(z) \in \singlevariablepolynomials{\finiteIntegerRing{\cnt}}{z}$.
From section \ref{Sec-polynomials-over-Zn}, it can be recalled that, $f(z) \in \invertibleRingElements{\cnt}$, for $z \in \finiteIntegerRing{\cnt}$, if and only if for every $i$, where $1 \leq i \leq r$, $f(z) \mod \basel{\primeInt}{i} \in \invertibleRingElements{\basel{\primeInt}{i}}$, for $z \in \finiteIntegerRing{\cnt}$.  

It may observed that if $a \in \nonzeroscalars$, $\cnt$ is the number of elements of $\nonzeroscalars$ and $l$ is a positive integer, then $a^{g(\basel{t}{1},\,\ldots,\, \basel{t}{l})} \in \nonzeroscalars$, for every $(\basel{t}{1},\,\ldots,\, \basel{t}{l}) \in \basel{\Integers^{l}}{\cnt}$ and any expression mapping $g(\basel{t}{1},\,\ldots,\, \basel{t}{l}) \in  \Expressions{\finiteIntegerRing{\cnt}}{t}{l}$.

\subsubsection{\label{Sec-invertible-square-matricies-with-multivariate-polynomial-entries}Invertible Square Matrices with Multivariate Mapping Entries}

For a positive integer $m$, a parametric $m \times m$ invertible square matrix is equivalent to a product of a permutation matrix, followed by a lower triangular matrix with nonzero diagonal entries, an upper triangular matrix with nonzero diagonal entries and finally by another permutation matrix, the four matrices being parametric and written from left to right in the product. Parametric permutation matrices can be constructed from a partition of unity. Let $\basel{g}{s}(\basel{z}{1},\,\ldots,\, \basel{z}{l})$, for $0 \leq s \leq m-1$, be a partition of unity of $\scalars^{l}$, which may not necessarily be strict. Let $\sigma \in \basel{\Integers}{m} \times \basel{\Integers}{m} \to \basel{\Integers}{m}$ be a mapping such that for each fixed $r \in \basel{\Integers}{m}$, $\sigma(r,\, \cdot)$ is a bijective mapping (permutation of indexes) from $\basel{\Integers}{m}$ into itself as a mapping of the second argument, and for each fixed $s \in \basel{\Integers}{m}$, $\sigma(\cdot,\, s)$ is a bijective mapping (permutation of indexes) from $\basel{\Integers}{m}$ into itself, as a mapping of the first argument. Then, the matrix with entries $\basel{g}{\sigma(i-1,\, j-1)}(\basel{z}{1},\,\ldots,\, \basel{z}{l})$ in the $i$-th row and $j$-th column, for $1 \leq i,\, j \leq m$, is a parametric permutation matrix. For an example of an index map $\sigma$ as discussed, let $\basel{f}{i}$ and $h$ be bijective mappings from $\basel{\Integers}{m}$ into itself, for $i \in \{1,\, 2\}$, and $\sigma(r,\, s) = h\bglb (\basel{f}{1}(r)+\basel{f}{2}(s)) \mod m\bgrb$. It can be easily checked that the mapping $\sigma$ is as required. Products and transposes of parametric permutation matrices are also parametric permutation matrices. 
\\

\noindent{\underline{\bf {Caution!}}} This paragraph is concerning an important restriction for parametric surjective mappings onto $\mygrp^{m}$ (and also for verification bijective mappings advertised in public key tables) in digital signature applications, when $\mygrp = \nonzeroscalars$, for a finite field $\scalars$, for exponential mappings. The parametric  lower and upper triangular matrices need to be chosen to be a parametric diagonal matrix with nonzero diagonal entries, {\em i.e.}, with entries that are multivariate mappings evaluating to invertible elements, for every assignment of values for the variables in their domains. The parametric permutation matrices are still permitted, in any case. The reason for this caution is the difficulty to deal with test-for-zero conditions. For overcoming this restriction, the exponential mappings need to be extended to  mappings that include $0 \in \scalars$ in their domains and co-domains, mapping $0$ to itself, but the test-for-zero conditions must be very carefully considered.   

\subsection{\label{Sec-nonparametric-univariate-bijective-mappings}Univariate Bijective Mappings without Parameters}

\subsubsection{Single Variable Permutation Mappings without Hashing} 

\noindent{\underline{\bf{Examples in} $\singlevariablepolynomials{\scalars}{x}$.}}~
Bijective mappings in  $\singlevariablepolynomials{\scalars}{x}$, also called  permutation polynomials, are extensively studied as Dickson polynomials [\cite{Dickson:1897}] in the literature. A comprehensive survey on Dickson polynomials can be found in [\cite{AG:1991},  \cite{GM:1994},   \cite{LMT:1993},  \cite{Mullen:2000} and  \cite{MN:1987}]. Some recent results are presented in [\cite{AAW:2008}, \cite{AGW:2009} and \cite{AW:2005}]. If $f(z) \in \singlevariablepolynomials{\scalars}{z}$
is a permutation polynomial, then, for every $a \in \scalars \backslash \{0\}$, 
$b \in \scalars$ and nonnegative integer $i$, the polynomial $af(z^{\primeInt^{i}})-b$
is a permutation polynomial. Some easy examples are described in the following.
\\

\begin{small}
\noindent{\bf{Examples.}}~   (A) ~ Let $\scalars$ be a finite dimensional
extension field of $\basel{\Integers}{\primeInt}$ of vector space dimension $n$. Any polynomial  $f(z) = \sum_{i = 1}^{n}\basel{a}{i}z^{\primeInt^{i-1}}$, where $\basel{a}{i} \in \scalars$, $1 \leq i \leq n$, that is an invertible linear operator from $\scalars$ onto $\scalars$, with
$\basel{\Integers}{\primeInt}$ as the field, is a permutation polynomial.~~
(B)~ Let $r$ be a positive integer divisor of $n$,
and $f(z) = z^{^{\primeInt^{r}}} - a z$,
where $a^{^{(\sum_{i = 1}^{n/r}\primeInt^{(i-1)r})}} \neq 1$.
Then, for every  $z \in \scalars \backslash \{0\}$,
$z^{^{(\primeInt^{r}-1)}}-a \neq 0$, since 
$z^{^{\primeInt^{n}}-1}  =  z^{^{(\primeInt^{r}-1)
\sum_{i = 1}^{n/r}\primeInt^{(i-1)r}}} = 1$,  and
therefore, the null space of $f(z)$, as a linear operator
from $\scalars$ into $\scalars$ with $\basel{\Integers}{\primeInt}$
as the field, is $\{0\}$. Thus, $f(z)$ is a permutation polynomial.~~
(C) ~Let $r$ be a positive integer relatively prime to $(\primeInt^{n}-1)$.
Then, the polynomial $f(z) = z^{r}$ is a permutation polynomial.
\\
\end{small}

\noindent\underline{{\bf{Examples in} $\singlevariablepolynomials{\finiteIntegerRing{\primeInt^{l}}}{x}$.}}~~ Let $l \in \PositiveIntegers$ and $\primeInt$ be a prime number. For any positive integer $n$, Dickson polynomials that are permutation polynomials, having nonvanishing derivatives over the finite field containing $\primeInt^{n}$ elements, are found in [\cite{AG:1991},  \cite{AAW:2008},  \cite{AGW:2009},   \cite{AW:2005},  \cite{GM:1994},  \cite{LMT:1993},  \cite{Mullen:2000} and  \cite{MN:1987}]. For a small prime number $\primeInt$, two methods for construction of permutation polynomials $f(x) \in \singlevariablepolynomials{\finiteIntegerField}{x}$, such that $f'(x) \not \equiv 0 \mod \primeInt$, are described below. As a set, $\finiteIntegerField$ is taken to be the set of integers $i$, where $0 \leq i \leq \primeInt-1$. For $\primeInt = 2$, since $x^{i+1} \equiv x^{i} \equiv x \mod 2$, for every $i \in \PositiveIntegers$ and $x \in \finiteIntegerRing{2^{l}}$, the only permutation polynomial mappings in $\singlevariablepolynomials{\finiteIntegerRing{2^{l}}}{x}$ are of the form $\basel{b}{0} + \sum_{i = 1}^{k} \basel{b}{i}x^{i}$, for some $k \in \PositiveIntegers$, $\basel{b}{i} \in \finiteIntegerRing{2^{l}}$, for $0 \leq i \leq k$, such that $\sum_{i = 1}^{k} \basel{b}{i} \equiv 1 \mod 2$, and, when $l \geq 2$, $\basel{b}{1} \equiv 1 \mod 2$ and $\sum_{i = 2}^{k} i\basel{b}{i} \equiv 0 \mod 2$, or equivalently,  $\basel{b}{1} \equiv 1 \mod 2$ and the number of indexes $j$, with $\basel{b}{2j+1} \equiv 1$ and $2 \leq 2j+1 \leq k$, is an even integer, for the condition $f'(x) \equiv 1 \mod 2$ to hold.  

Now, let $\primeInt \geq 3$ be a small prime number, such that the computations below are not difficult for implementation. Let {\small{$\basel{\ell}{i}(x) = \bgls \prod_{{\tiny{\begin{array}{c} j = 0\\ j \neq i \end{array}}}}^{\primeInt-1} (i-j)\bgrs^{-1} \cdot
 \prod_{{\tiny{\begin{array}{c} j = 0\\ j \neq i \end{array}}}}^{\primeInt-1} (x-j) = 
- \prod_{{\tiny{\begin{array}{c} j = 0\\ j \neq i \end{array}}}}^{\primeInt-1} (x-j) $},} since
$\finiteIntegerRing{\primeInt}$ is the solution set for $x$ in the polynomial equation
$x^{\primeInt-1}-1 = 0$, and hence,
{\small{$\bgls \prod_{{\tiny{\begin{array}{c} j = 0\\ j \neq i \end{array}}}}^{\primeInt-1} (i-j)\bgrs^{-1}$
$ = $
$-1$},}
 for $i \in \finiteIntegerField$.
 Now, {\small{$\basel{\ell'}{i}(x) = - \sum_{{\tiny{\begin{array}{c} j = 0\\ j \neq i \end{array}}}}^{\primeInt-1}   \prod_{{\tiny{\begin{array}{c} k = 0\\ k \not \in \{i,\, j\} \end{array}}}}^{\primeInt-1} (x-k)$, for $i \in \finiteIntegerField$}~,}~ which implies that {\small{ $\basel{\ell'}{i}(j) =  - \prod_{{\tiny{\begin{array}{c} k = 0\\ k \not \in \{i,\, j\} \end{array}}}}^{\primeInt-1} (j-k) ~ = ~ (j-i)^{-1}$~,~~ for $j \neq i$ and $j \in \finiteIntegerField$},} and  {\small{$\basel{\ell'}{i}(i) =   - \sum_{{\tiny{\begin{array}{c} j = 0\\ j \neq i \end{array}}}}^{\primeInt-1}   \prod_{{\tiny{\begin{array}{c} k = 0\\ k \not \in \{i,\, j\} \end{array}}}}^{\primeInt-1} (i-k) ~ = ~ \sum_{{\tiny{\begin{array}{c} j = 0\\ j \neq i \end{array}}}}^{\primeInt-1}  (i-j)^{-1} = 0~$},}  for $i \in \finiteIntegerField$, since $\primeInt \geq 3$. For a fixed permutation sequence $\{\basel{a}{i} \in \finiteIntegerField \,:\, 0 \leq i \leq \primeInt-1\}$ of  $\finiteIntegerField$, either of the two procedures described below constructs a permutation polynomial $f(x) \in \singlevariablepolynomials{\finiteIntegerField}{x}$, such that $f(i) = \basel{a}{i}$ and $f'(i) \not \equiv 0 \mod \primeInt$, for $i \in \finiteIntegerField$.
\\

  \noindent
{\small{\fbox{\sf{Method 1}}}}~~
Let $\sum_{i = 0}^{\primeInt-1} \basel{a}{i} \basel{\ell}{i}(x) = \basel{b}{0} + \sum_{i = 1}^{\primeInt-1} \basel{b}{i} x^{i}$, for some $\basel{b}{i} \in \finiteIntegerField$, for $0 \leq i \leq \primeInt-1$, and let $g(x) = \basel{c}{1} + \sum_{i=2}^{\primeInt-1} \basel{c}{i} x^{i-1}$, for some $\basel{c}{i} \in \finiteIntegerField$, for $1 \leq i \leq \primeInt-1$, be such that $g(x) \not \equiv 0 \mod \primeInt$, for every $x \in \finiteIntegerField$. Let $\basel{\rho}{i} = i^{-1} \basel{c}{i}$ and $\basel{\sigma}{i} = \basel{b}{i}-\basel{\rho}{i}$, for $1 \leq i \leq \primeInt-1$. Let $f(x) = \basel{b}{0} + \sum_{i = 1}^{\primeInt-1} (\basel{\rho}{i} x^{i} + \basel{\sigma}{i}x^{i\primeInt})$. Then, $f(x) \equiv \basel{b}{0} + \sum_{i = 1}^{\primeInt-1} \basel{b}{i} x^{i} \mod \primeInt$, for every $x \in \finiteIntegerField$, and $f'(x) \equiv \basel{\rho}{1} + \sum_{i = 2}^{\primeInt-1} i \basel{\rho}{i} x^{i-1} \equiv  \basel{c}{1} + \sum_{i = 2}^{\primeInt-1}  \basel{c}{i} x^{i-1} \mod \primeInt$, for every $x \in \finiteIntegerField$, satisfying the stated requirement. In this method, $\deg\bglb f(x) \bgrb$ can be as high as $(\primeInt-1)\primeInt$. In the next method,  $\deg\bglb f(x) \bgrb$ is at most $(2\primeInt-2)$.
\\

  \noindent
{\small{\fbox{\sf{Method 2}}}}~~
Let $\basel{b}{i}, \, \basel{c}{i},\, \sigma \in \finiteIntegerField$, for $ 0 \leq i \leq \primeInt-1$, be such that $\basel{b}{0} = \basel{a}{0}$ and $\basel{b}{j}+\basel{c}{j} = \basel{a}{j}$, for $1 \leq j \leq \primeInt-1$, and let $f(x) =  \sum_{i = 0}^{\primeInt-1} (\basel{b}{i} + x^{\primeInt-1}\basel{c}{i} - \sigma i)\basel{\ell}{i}(x) + \sigma x^{\primeInt}$. It can be immediately verified that $f(i) \equiv \basel{a}{i} \mod \primeInt$, for $0 \leq i \leq \primeInt-1$, and $f'(x) =  \sum_{i = 0}^{\primeInt-1} (\basel{b}{i}  +  x^{\primeInt-1} \basel{c}{i} - \sigma i) \basel{\ell'}{i}(x) + \primeInt \sigma x^{\primeInt-1} + (\primeInt-1)x^{\primeInt-2} \sum_{i = 0}^{\primeInt-1} \basel{c}{i}\basel{\ell}{i}(x)$, where $\primeInt \geq 3$. Thus, the parameters $\basel{c}{0}$, $\sigma$, $\basel{b}{j}$ and $\basel{c}{j}$, for $1 \leq j \leq \primeInt-1$, need to be chosen such that $f'(x) \not \equiv 0 \mod \primeInt$, for all $x \in \finiteIntegerField$. Now, $f(x)+ \sigma x =  \sum_{i = 0}^{\primeInt-1} (\basel{b}{i} + \basel{c}{i} x^{\primeInt-1}) \basel{\ell}{i}(x) + \sigma x^{\primeInt}$, and $f'(x)+\sigma =  \sum_{i = 0}^{\primeInt-1} (\basel{b}{i} + \basel{c}{i} x^{\primeInt-1}) \basel{\ell'}{i}(x) + \primeInt \sigma x^{\primeInt-1} + (\primeInt-1)x^{\primeInt-2} \sum_{i = 0}^{\primeInt-1} \basel{c}{i}\basel{\ell}{i}(x)$.  Thus,  $f'(0)+\sigma \equiv  - \sum_{i = 1}^{\primeInt-1} i^{-1} \basel{b}{i} \mod \primeInt$ and $f'(j) +\sigma \equiv  \sum_{{\tiny{\begin{array}{c} i = 0\\ i \neq j \end{array}}}}^{\primeInt-1} \basel{a}{i} (j-i)^{-1} + \basel{c}{0} j^{-1} - j^{-1} \basel{c}{j} \mod \primeInt$, for $1 \leq j \leq \primeInt-1$, which implies that every element in the sequence of numbers $(f'(i)+\sigma) \mod \primeInt$, for $0 \leq i \leq \primeInt-1$, is independent of the choice of $\sigma$, and the condition that $f'(i) \not \equiv 0 \mod \primeInt$, for $0 \leq i \leq \primeInt-1$, is equivalent to that $\sigma \not \in \{(f'(i)+\sigma) \mod \primeInt\,:\, 0 \leq i \leq \primeInt-1\}$. For $\primeInt \geq 3$, $\sum_{i = 0}^{\primeInt-1} i \equiv \sum_{i = 0}^{\primeInt-1} 1 \equiv 0 \mod \primeInt$, and since $\finiteIntegerField$ is the splitting field of the polynomial $x^{\primeInt} - x = \prod_{i = 0}^{\primeInt-1}(x-i)$, the elementary symmetric polynomials $\basel{s}{r}(\basel{t}{1},\, \basel{t}{2},\, \ldots,\, \basel{t}{n})$, which are homogeneous of degree $r$ in $n$ variables, for the particular instances of parameters $n = \primeInt$ and $\basel{t}{i}= i-1$, for $1 \leq i \leq \primeInt$, as defined in [\cite{Lang:2002}], are all congruent to $0 \mod \primeInt$, for $1 \leq r \leq \primeInt-2$. Thus, $\sum_{i = 0}^{\primeInt-1} i^{r} \equiv \sum_{i = 0}^{\primeInt-1} 1 \equiv 0 \mod \primeInt$, for $r \in \PositiveIntegers$, $1 \leq r \leq \primeInt-2$ and $\primeInt \geq 3$, which implies that for a nonzero polynomial $g(x) \in \singlevariablepolynomials{\finiteIntegerField}{x}$ of degree at most $\primeInt-2$, $\sum_{i = 0}^{\primeInt-1} g(i) \equiv 0 \mod \primeInt$. Now, $\primeInt \sum_{i = 0}^{\primeInt-1} i^{\primeInt-1} \equiv 0 \mod \primeInt$, and, for $l \in \PositiveIntegers$, such that $\primeInt+1 \le l \leq 2\primeInt-2$, $l\sum_{i = 0}^{\primeInt-1} i^{l-1} \equiv l\sum_{i = 0}^{\primeInt-1} i^{l-1-(\primeInt-1)} \equiv  l\sum_{i = 0}^{\primeInt-1} i^{l-\primeInt} \equiv 0 \mod \primeInt$, since $1 \leq l-\primeInt \leq \primeInt-2$. Thus, for a nonzero polynomial $h(x) \in \singlevariablepolynomials{\finiteIntegerField}{x}$ of degree at most $2\primeInt-2$, $\sum_{i = 0}^{\primeInt-1} h'(i) \equiv 0 \mod \primeInt$. The coefficients $\basel{c}{i}$, for $0 \leq i \leq \primeInt-1$, must be so chosen that the additional requirement that $f(x)+\sigma x$ is a polynomial of degree at most $2\primeInt-2$ can also be fulfilled. Now, let $\basel{\lambda}{i} \in \finiteIntegerField$, for $0 \leq i \leq \primeInt-1$, be chosen, such that the cardinality of the set $\Lambda = \{\basel{\lambda}{i}\,:\, 0 \leq i \leq \primeInt-1\}$ is at most $\primeInt-1$ and $\sum_{i = 0}^{\primeInt-1} \basel{\lambda}{i} = 0$. Then, $\basel{c}{j}-\basel{c}{0}$ are found from the condition $f'(j)+\sigma = \sum_{{\tiny{\begin{array}{c} i = 0\\ i \neq j \end{array}}}}^{\primeInt-1} \basel{a}{i} (j-i)^{-1}  - j^{-1} (\basel{c}{j}-\basel{c}{0}) = \basel{\lambda}{j}$, for $1 \leq j \leq \primeInt-1$, and hence, $ f'(0)+\sigma =  - \sum_{i = 1}^{\primeInt-1} i^{-1} \basel{b}{i}  = \basel{\lambda}{0}$, for all choices of $\basel{c}{0}$. Now, let $\sigma$ be chosen from $\finiteIntegerField \backslash \Lambda$, where the latter set is nonempty, since the cardinality of $\Lambda$ is at most $\primeInt-1$, by the choices of $\basel{\lambda}{i}$, for $0 \leq i \leq \primeInt-1$. Finally, $\basel{c}{0}$ is chosen, and $\basel{b}{j}$ and $\basel{c}{j}$, for $1 \leq j \leq \primeInt-1$, are determined by the aforementioned conditions.
 
 For a small prime number $\primeInt$, positive integers $l$ and $r$, such that $l \geq 2$ and $1 \leq r \leq l$, a bijective mapping $f(x) \in \singlevariablepolynomials{\finiteIntegerRing{\primeInt^{l}}}{x}$ and $y \in \finiteIntegerRing{\primeInt^{l}}$, the following procedure computes $\basel{x}{r} \in \finiteIntegerRing{\primeInt^{r}}$, such that $\basel{f}{r}(\basel{x}{r}) \equiv y \mod \primeInt^{r}$, assuming $\basel{x}{1} \in \finiteIntegerField$ is known, such that $\basel{f}{1}(\basel{x}{1}) \equiv y \mod \primeInt$, where $\basel{f}{r}(x) =  f(x) \mod \primeInt^{r}$, applying the $\mod \primeInt^{r} ~ $ operation only to the coefficients. Let $2 \leq r \leq l$, where $l \geq 2$, $s \in \PositiveIntegers$ be such that $\left \lceil \frac{r}{2} \right \rceil \leq s \leq r-1$ and $\basel{y}{r} = y \mod \primeInt^{r} \in \finiteIntegerRing{\primeInt^{r}}$, and $\basel{x}{s} = \basel{f^{-1}}{s}(\basel{y}{r} \mod \primeInt^{s}) \in \finiteIntegerRing{\primeInt^{s}}$ has been computed. Let $\basel{\hat{x}}{s} \in \finiteIntegerRing{\primeInt^{r}}$ be such that $\basel{\hat{x}}{s} \equiv \basel{x}{s} \mod \primeInt^{s}$. Since $\basel{f}{r}(\basel{\hat{x}}{s}) \equiv \basel{y}{r} \mod \primeInt^{s}$, it follows that $\basel{f}{r}(\basel{\hat{x}}{s}) = \basel{y}{r} + \primeInt^{s} \basel{g}{r,\,s }(\basel{\hat{x}}{s},\, \basel{y}{r})$, for some mapping $\basel{g}{r,\, s}(\basel{\hat{x}}{s},\, \basel{y}{r})$, and therefore, $\basel{f}{r}\bglb {\small{\basel{\hat{x}}{s} +  [ \basel{f'}{r}(\basel{\hat{x}}{s})]^{-1} \cdot [ \basel{y}{r} - \basel{f}{r}(\basel{\hat{x}}{s}) ]}} \bgrb \equiv \basel{f}{r}({\small{\basel{\hat{x}}{s}}}) +  \basel{f'}{r}({\small{\basel{\hat{x}}{s}}}) \cdot \bgls \basel{f'}{r}({\small{\basel{\hat{x}}{s}}})\bgrs^{-1} \cdot \bgls \basel{y}{r} - \basel{f}{r}({\small{\basel{\hat{x}}{s}}}) \bgrs \equiv  \basel{f}{r}({\small{\basel{\hat{x}}{s}}}) +  \bgls \basel{y}{r} - \basel{f}{r}({\small{\basel{\hat{x}}{s}}}) \bgrs  \equiv \basel{y}{r} \mod \primeInt^{r}$. Thus, $\basel{f^{-1}}{r}(\basel{y}{r}) = \basel{\hat{x}}{s} +  \bgls \basel{f'}{r}(\basel{\hat{x}}{s})\bgrs^{-1} \cdot \bgls \basel{y}{r} - \basel{f}{r}(\basel{\hat{x}}{s}) \bgrs \mod \primeInt^{r}$. If $r = l$, then the $f^{-1}(y)$ is just computed for $y \in \finiteIntegerRing{\primeInt^{l}}$, and the procedure can be stopped; otherwise, the previous steps are repeated, replacing the current value of $r$ by $\min\{2r,\, l\}$. 
\\
 
\noindent{\underline{\bf{Examples in} $\singlevariableexpressions{\scalars}{z}$.}}~
 Let $\scalars$ be a finite field of $\primeInt^{n}$ elements, for some prime number $\primeInt$ and $n \in \PositiveIntegers$, such that $\primeInt^{n} \geq 3$, and let $\cnt = \primeInt^{n}-1$. Let $s, \, t \in \PositiveIntegers$ be such that $\gcd(s,\, t) = 1$, $st = \cnt$ and   $2 \leq s,\, t \leq \cnt-1$, and let $\basel{H}{t} = \{x^{t} = 1 \, : \, x \in \nonzeroscalars\}$. Let $f(x)\in \singlevariablepolynomials{\Integers}{x}$ be such that $f(x) \mod t$ yields a polynomial mapping from $\finiteIntegerRing{t}$ onto itself. It may be recalled that, as a set, $\basel{\Integers}{t}$ is assumed to consist of integers $i$, where $0 \leq i \leq t-1$. Let $a$ be a primitive element in $\nonzeroscalars$. Now, for $x \in \basel{H}{t}$, since $x^{t} = 1$, applying $\basel{\log}{a}$ operation on both sides, $t \basel{\log}{a} x = 0 \mod \cnt$, which implies that $\basel{\log}{a} x$ is an integer multiple of $s$, for every $x \in \basel{H}{t}$, and, since the cyclic subgroup generated by $a^{s}$ is $\basel{H}{t}$, it follows that $\basel{\log}{a}$ is a bijective mapping of $\basel{H}{t}$ onto $s \cdot \finiteIntegerRing{\cnt}$. Now, $f(\basel{\log}{a}(x)) \mod \cnt$, for $x \in \basel{H}{t}$, is an injective mapping, when restricted to $\basel{H}{t}$, which can be modified appropriately, by changing its constant term, if necessary, to obtain a polynomial $g$, which results in a bijective mapping from $s\cdot \finiteIntegerRing{\cnt}$ into itself, with respect to $\mod \cnt$ operation.  Then, the mapping $\eta(x) = a^{g(\basel{\log}{a} x)}$, for $x \in \nonzeroscalars$, is such that its restriction to $\basel{H}{t}$ is a bijective mapping from $\basel{H}{t}$ onto itself.

\subsubsection{Hybrid Single Variable Permutation Mappings with Hashing}

\noindent
{\small{\fbox{\sf{Method 1}}}}~~
Let $\basel{\ell}{i}(x) \in \singlevariablepolynomials{\scalars}{x}$, $1 \leq i \leq k$, where $k \in \PositiveIntegers$, $k \geq 2$, be indicator functions of a partition $\{\basel{S}{i}\,:\, 1 \leq i \leq k\}$  of $\scalars$. Let $\sigma$ be a permutation on $\{1,\, \ldots,\, k\}$, such that the set cardinalities of $\basel{S}{i}$ and $\basel{S}{\sigma(i)}$ are equal, for $1 \leq i \leq k$. Let $\basel{g}{i}$ be a mapping from $\scalars$ into $\scalars$, such that $\basel{g}{i}\bglb \basel{S}{i} \bgrb = \basel{S}{\sigma(i)}$, for $1 \leq i \leq k$. Thus, $\basel{g}{i}$ is one-to-one when restricted to $\basel{S}{i}$, for $1 \leq i \leq k$. Let $\eta(x) \in \singlevariablepolynomials{\scalars}{x}$ be a permutation polynomial, and $\chi(x) = \sum_{i = 1}^{k} \basel{\ell}{i}(x) \eta({\small{\basel{g}{i}(x)}})$. Then, $\chi(\scalars) = \bigcup_{i = 1}^{k} \eta\bglb \basel{g}{i}(\basel{S}{i})\bgrb = \bigcup_{i = 1}^{k} \eta\bglb \basel{S}{\sigma(i)} \bgrb$, and  since $\{\basel{S}{\sigma(i)}\,:\, 1 \leq i \leq k\}$ is a partition of $\scalars$, $\chi(x)$ is a surjective (hence bijective) polynomial from $\scalars$ onto $\scalars$. For inverting $\chi(x) = y$, for fixed $y \in \scalars$, let $\xi = \eta^{-1}(y)$. Now, there exists exactly one index $i$, where $1 \leq i \leq k$, such that $\xi \in \basel{S}{\sigma(i)} = \basel{g}{i}\bglb\basel{S}{i}\bgrb$, and therefore, the unique element $x \in \basel{S}{i}$, such that $x  = \basel{g^{-1}}{i}(\xi)$, satisfies $\chi(x) = y$. If $\basel{f}{i}$, for $1 \leq i \leq k$, are mappings from $\scalars$ into $\scalars$, such that $\basel{f}{i}(\basel{g}{i}(x)) = x$, for $x \in \basel{S}{i}$, then $\chi^{-1}(y) = \sum_{i = 1}^{k} \basel{\ell}{\sigma(i)} \bglb \eta^{-1}(y)\bgrb \basel{f}{i}\bglb \eta^{-1}(y) \bgrb$, for $y \in \scalars$. The case of bijective mappings in $\singlevariableexpressions{\scalars}{x}$ can be similarly discussed. In the following examples, the corresponding examples in section \ref{Sec-partition-of-unity} are  revisited.
\\

\begin{small}
\noindent{\bf{Examples.}}   (A) ~ Let $T(x) = \sum_{i = 1}^{n} \basel{a}{i}x^{\primeInt^{i-1}}$, $\basel{a}{i} \in \scalars$, $1 \leq i \leq n$,  be of rank $t$, where  $t$ is a small positive integer, such as $t \in \{1,\, 2\}$,  as  described in the first example in section \ref{Sec-partition-of-unity} and let $V = \{x \in \scalars \, : \, T(x) = 0\}$. Then, there exist $k = \primeInt^{t}$ representative elements $\basel{b}{i} \in \scalars$, $1 \leq i \leq k$, such that $\{T(\basel{b}{i})\,:\,  1 \leq i \leq k\} = T(\scalars)$, and  $\basel{S}{i} =   V + \basel{b}{i} = \{x + \basel{b}{i} \, : \, x \in V\}$,  $1 \leq i \leq k$.  Let $\basel{f}{i}(x) = \basel{c}{i,\, 0}+ \sum_{i = 1}^{n} \basel{c}{i,\, j} x^{\primeInt^{j-1}}$, where $\basel{c}{i,\, j},\, x \in \scalars$, $0 \leq j \leq n$, be such that $V \subseteq \basel{f}{i} (V)$, for $1 \leq  i \leq k$.   Thus,  in the notation of the above discussion, the permutation polynomial $\basel{f}{i} (x)-\basel{b}{i}+\basel{b}{\sigma(i)}$ can be chosen to be $\basel{g}{i}(x)$, for $x \in \scalars$ and $1 \leq i \leq k$. ~~(B) ~ Let $f(z) = z^{t}$,  where $t$ is a large positive integer dividing $\primeInt^{n}-1$, as described in the second example of section \ref{Sec-partition-of-unity}.  Let $\basel{a}{1} = 0$ and $\basel{a}{i} \in \nonzeroscalars$, for $2 \leq i \leq k$, where $k = 1+\frac{(\primeInt^{n}-1)}{t}$, be such that $\{f(\basel{a}{i}) \,:\, 1 \leq i \leq k\}$ is the codomain of $f$. Let $\sigma$ be a permutation on $\{1,\, \ldots,\, k\}$, such that $\sigma(1) = 1$, and let $\basel{H}{t} = \{y \in \scalars \,:\, y^{t} = 1\}$. Then, $\basel{S}{i} = \basel{a}{i}\basel{H}{t} = \{\basel{a}{i}v \,:\, v \in \basel{H}{t}\}$, for $1 \leq i \leq k$. Let $\basel{h}{i}(x)$, $x \in \basel{H}{t}$, be a bijective mapping discussed in the previous section, for $2 \leq i \leq k$. Thus, representing elements $\basel{c}{i} \in \nonzeroscalars$ can be found easily, such that the mapping $\basel{g}{i}(x) = \basel{c}{i} \basel{h}{i}(\basel{a^{-1}}{i} x)$ satisfies $\basel{g}{i}\bglb \basel{S}{i} \bgrb = \basel{S}{\sigma(i)}$, for $x \in \basel{S}{i}$ and $2 \leq i \leq k$.
\\

\end{small}

\noindent
{\small{\fbox{\sf{Method 2}}}}~~
Let $\mygrp$ be $\nonzeroscalars$ or $\scalars$. Let $k,\, \rho \in \PositiveIntegers$, such that $2 \leq k \leq \rho$. Let $\basel{f}{i}$ be bijective mappings from $\mygrp$ into itself, for $1 \leq i \leq \rho$, and $h$ be a mapping from $\mygrp$ into itself, such that $h\bglb {\small{\basel{f}{i}(x)}} \bgrb = h\bglb {\small{\basel{f}{j}(x)}} \bgrb $, for $x \in \mygrp$ and $1 \leq i,\, j \leq \rho$.  Let $\sigma$ be a permutation on $\{1,\, \ldots, \, \rho\}$, and $\{\basel{S}{i}\, : \, 1 \leq i \leq k \}$ be a partition of $\scalars$, and let $\basel{\ell}{i}(x)$, $x \in \scalars$, be the indicator function of $\basel{S}{i}$, for $1 \leq i \leq k$. Let $\eta$ be a bijective mapping from $\mygrp$ into $\mygrp$, and $\zeta(x) = \sum_{i = 1}^{k} \basel{\ell}{i}\bglb h(x)\bgrb \eta\bglb\basel{f}{\sigma(i)}(x)\bgrb$, for $x \in \mygrp$. Let $x, \, y \in \mygrp$ be such that $\zeta(x) = \zeta(y)$, and let $i,\, j \in \{1,\, \ldots, \, k\}$ be such that $\basel{\ell}{i}(h(x)) = 1$ and $\basel{\ell}{j}(h(y)) = 1$. Then, $\eta\bglb\basel{f}{\sigma(i)}(x)\bgrb = \eta\bglb \basel{f}{\sigma(j)}(y)\bgrb$, and  since $\eta$ is bijective, it follows that $\basel{f}{\sigma(i)}(x) = \basel{f}{\sigma(j)}(y)$. Now, since $h\bglb {\small{\basel{f}{i}(x)}} \bgrb = h\bglb {\small{\basel{f}{j}(x)}} \bgrb $, for $x \in \mygrp$ and $1 \leq i,\, j \leq \rho$, and $\sigma$ is a permutation on the set $\{1,\, \ldots,\, \rho\}$, it follows that $h(x) = h(y)$, $\sigma(i) = \sigma(j)$ and $i = j$, and therefore, $x = y$. Thus, $\zeta^{-1}(y) = \sum_{i = 1}^{k} \basel{\ell}{i}\bglb h({\small{\eta^{-1}(y)}})\bgrb \basel{f^{-1}}{\sigma(i)}\bglb {\small{\eta^{-1}(y)}} \bgrb$, for $y \in \mygrp$.
\\

\begin{small}
\noindent{\bf{Examples.}} (A)~ Let ($i$) $f$ be a bijective mapping from $\mygrp$ into itself, such that the cyclic group generated by it, as a subgroup of bijective mappings from $\mygrp$ into $\mygrp$, with composition as the group operation, is of small order $\rho \geq 2$, ($ii$) $g\, : \, \mygrp^{\rho} \rightarrow \scalars$ is a symmetric function, which can be  an expression in $\Expressions{\scalars}{z}{\rho}$, symmetric in all the $\rho$ variables, ($iii$) $\basel{f}{0}(x) = x$ and  $\basel{f}{i}(x) = f\bglb\basel{f}{i-1}(x)\bgrb$, for $1 \leq i \leq \rho$,  and ($iv$) $h(x) = g\bglb x,\, \basel{f}{1}(x),\, \ldots, \, \basel{f}{\rho-1}(x)\bgrb$, for $x \in \mygrp$. Then, $\basel{f}{\rho}(x) = x$ and, since $h\bglb f(x)\bgrb = h(x)$, for $x \in \mygrp$,  it follows that $h\bglb {\small{\basel{f}{i}(x)}} \bgrb = h\bglb {\small{\basel{f}{j}(x)}} \bgrb $, for $x \in \mygrp$ and $1 \leq i,\, j \leq \rho$. If $\mygrp = \nonzeroscalars$, then it is interesting to choose $f(x) = a^{\phi(\basel{\log}{a}x)}$, for $x \in \nonzeroscalars$ and some primitive element $a \in \nonzeroscalars$. However, it is important to choose $f$ such that $\rho$ is a small positive integer. ~~(B)~ Let $\mygrp = \nonzeroscalars$ and $s,\, t,\, v \in \PositiveIntegers$ be such that $\gcd(s,\, t) = 1$, $st =\cnt$, $2 \leq s,\, t \leq \cnt-1$, $sv = 1 \mod t$ and $s$ large. Let $\phi\,:\, \basel{\Integers}{\cnt} \rightarrow  \basel{\Integers}{\cnt}$ be a polynomial mapping such that $sv\phi(y)$ is a bijective mapping from $sv\basel{\Integers}{\cnt}$ into itself and the order of the cyclic group generated by $sv\phi(y)$ as a subgroup of the group of bijective mappings from $sv\basel{\Integers}{\cnt}$ into itself is a small positive integer $\rho$. Now, let $\pi(x) = a^{sv\phi(\basel{\log}{a} x)}$, for $x \in \nonzeroscalars$, where $a$ is a primitive element in $\nonzeroscalars$. Then, $\pi\bglb \nonzeroscalars \bgrb =  \pi\bglb \basel{H}{t}\bgrb = \basel{H}{t}$, where $\basel{H}{t} = \{x \in \nonzeroscalars\,:\, x^{t} = 1\}$. Let $\basel{\pi}{1} = \pi$ and $\basel{\pi}{i+1} = \basel{\pi}{i}(\pi)$, for $i \in \PositiveIntegers$. Then, $\basel{\pi}{\rho+1}(x) = \basel{\pi}{1}(x)$, for $x \in \basel{H}{t}$. Let $\basel{f}{i}$ be bijective mappings from $\nonzeroscalars$ into itself, such that the restriction of $\basel{f}{i}$ to $\basel{H}{t}$ is $\basel{\pi}{i}$, for $1 \leq i \leq \rho$, and $g$ be the symmetric mapping as in the previous example and $h(x) = g\bglb\basel{\pi}{1}(x),\, \ldots, \basel{\pi}{\rho}(x)\bgrb$. It can be easily checked that  $h\bglb {\small{\basel{f}{i}(x)}} \bgrb = h\bglb {\small{\basel{f}{j}(x)}} \bgrb $, for $x \in \mygrp$ and $1 \leq i,\, j \leq \rho$.
\\

\end{small}
  
\subsection{\label{Sec-nonparametric-multivariate-injective-mappings}Multivariate Injective Mappings without Parameters}
 
\subsubsection{Multivariate Injective Mappings from $\mygrp^{m}$ into $\eltSet^{m}$}
In this subsection, an iterative algorithm to construct a multivariate bijective mapping from $\mygrp^{m}$ into $\eltSet^{m}$, for $m\in \PositiveIntegers$, is described. The algorithm utilises parametric univariate bijective mappings discussed in the previous sections. In later subsections, some variations involving hashing are described.

\begin{enumerate}
\item Let $\basel{f}{i}\,:\, \mygrp \rightarrow \mygrp$ and $\basel{g}{i}\,:\, \eltSet \rightarrow \eltSet$,  for $1 \leq i \leq m$, be bijective mappings. 

\item Let $\basel{h}{i}(\basel{z}{1},\, \ldots, \, \basel{z}{m-1};\, x)$ be parametric injective mappings from $\mygrp$ into $\eltSet$, for $1 \leq i \leq m$, $x \in \mygrp$ and $\basel{z}{1},\, \ldots, \, \basel{z}{m-1} \in \eltSet$ being parameters, constructed, for example, as described in section \ref{sec-parametrization-of-permutation-polynomials}.

\item Let  $\basel{\zeta}{i}(\xx) = \basel{h}{i}\bglb \basel{\zeta}{i+1}(\xx),\, \ldots,\,
 \basel{\zeta}{m}(\xx),\, \basel{x}{1},\, \ldots,\, \basel{x}{ i-1}\, ; ~
\basel{f}{i}(\basel{x}{i})\bgrb$ and  $\basel{\eta}{i}(\xx) = \basel{g}{i}\bglb \basel{\zeta}{i}(\xx) \bgrb$, for $\xx = (\basel{x}{1},\, \ldots,\, \basel{x}{m}) \in \mygrp^{m}$ and $1\leq i \leq m$. Let $\eta(\xx) = (\basel{\eta}{1}(\xx),\, \ldots,\, \basel{\eta}{m}(\xx))$.

\end{enumerate}
\noindent
For finding $\xx =(\basel{x}{1},\,\ldots,\, \basel{x}{m}) \in \mygrp^{m}$, such that $\eta(\xx) = \yy$,
for any fixed $\yy = (\basel{y}{1},\,\ldots,\, \basel{y}{m})$
$ \in $
$\eltSet^{m}$,  let {\small{$\basel{\epsilon}{i} = \basel{g^{-1}}{i}(\basel{y}{i})$}} and {\small{$\basel{\delta}{i} = \basel{h^{-1}}{i} (\basel{\epsilon}{i+1},\, \ldots, \, \basel{\epsilon}{m},\, \basel{x}{1},\, \ldots,\, \basel{x}{i-1};\, \basel{\epsilon}{i})$},} for $1 \leq i \leq m$.  Then, $\basel{x}{i} = \basel{f^{-1}}{i} (\basel{\delta}{i})$, for $1 \leq i \leq m$.  Now, for $\eltSet = \scalars$ and $\mygrp = \nonzeroscalars$, if $\basel{g}{i}$ and $\basel{h}{i}$,  for $1 \leq i \leq m$, are bijective mappings and parametric bijective mappings, respectively, from $\nonzeroscalars$ into $\nonzeroscalars$, then the above procedure can be applied to obtain multivariate bijective mappings from $\mygrp^{m}$ into $\mygrp^{m}$. These mappings are required in appealing for a security that is immune to threats resulting from Gr\"{o}bner basis analysis. It can be observed that one level of exponentiation suffices for the purpose. 
\\

 A one-to-one mapping from $\mygrp^{m}$ into $\eltSet^{n}$, where $m$ and $n$ are positive integers, with $m \leq n$, and $\mygrp$ is a subset of a finite field $\scalars$, is obtained as follows: for a carefully chosen bijective mapping $P(\yy)$ from $\mygrp^{n}$ into $\eltSet^{n}$ and hashing keys $\basel{f}{i}(\xx)$, for $\xx = (\basel{x}{1},\, \ldots,\, \basel{x}{m}) \in \mygrp^{m}$ and $1 \leq i \leq n-m$, the argument vector $(\basel{f}{1}(\xx),\, \ldots,\, \basel{f}{n-m}(\xx),\,  \basel{x}{1},\, \ldots,\, \basel{x}{m})$ is substituted for $\yy \in \mygrp^{n}$ in $P(\yy)$. Thus, $Q(\xx) = P(\basel{f}{1}(\xx),\, \ldots,\, \basel{f}{n-m}(\xx),\,  \basel{x}{1},\, \ldots,\, \basel{x}{m})$ is a generic multivariate one-to-one mapping from $\mygrp^{m}$ into $\mygrp^{n}$.

\subsubsection{Hybrid Multivariate Injective Mappings with Hashing}
For Method 1 of the previous subsection, in the first example, in place of $T(x)$, $x \in \scalars$,  $T\bglb\alpha(\xx)\bgrb$, $\xx \in \scalars^{m}$, and in the second example, in place of $f(z)$, $z \in \scalars$, $f\bglb \beta(\xx)\bgrb$, $\xx \in \scalars^{m}$, are chosen, where $\alpha \,: \, \scalars^{m} \rightarrow \scalars$ is a non constant affine mapping in the first example, and $\beta(\xx) = c \prod_{i = 1}^{m}\basel{x^{\basel{s}{i}}}{i}$, for some nonnegative integers  $\basel{s}{i}$, which, when positive, are relatively prime to $\primeInt^{n}-1$, and, when zero, for the corresponding subscript index $i$, the variable $\basel{x}{i}$ does not occur in the product, for $1 \leq i \leq m$, such that $\beta(\xx)$ is nonconstant, in the second example. Similarly, Method 2 hashing of the previous subsection can also be  extended to multivariate mappings, replacing $x$ with $\xx$, and choosing $\Phi(\yy) = (\basel{\phi}{1}(\yy),\, \ldots,\, \basel{\phi}{m}(\yy)\bgrb$ to be a bijective mapping from $\basel{\Integers^{m}}{\cnt}$ into itself in place of $\phi(y)$. It can be observed that $g$ can also be chosen to depend only on a few scalar components from each vector, while maintaining symmetry in all its vector parameters, with each vector consisting of $m$ scalars components. In the first example of Method 2 hashing of the previous section, if $h(\xx)$ is a symmetric mapping in its $m$ components, then $f(\xx)$ can be chosen to be a permutation of components of $\xx$, independent of order $\rho$ of the cyclic group generated by $f$, with respect to composition operation.

\section{\label{Sec-PKC-and-DS}Public Key Cryptography and Digital Signature}

  Let the number of elements in the plain message (or plain signature message) be $\mu$, and the number of elements in the encrypted message (or encrypted signature message) be $\nu$, where $\mu, \, \nu \in \PositiveIntegers$ and $\mu \leq \nu$.  Let $\eltSet$ be $\scalars$ or $\finiteIntegerRing{\cnt}$, and $\mygrp \subseteq \eltSet$ be the set from which plain message elements are sampled. If the number of plain and encrypted (or plain and signed) messages are the same, then a multivariate bijective mapping $P\,:\, \mygrp^{\mu} \rightarrow \mygrp^{\mu}$ is chosen and advertised in the public key lookup table $\LookupTable$, while $P^{-1}$ is saved in the back substitution table $\BackSubstitutionTable$. Let $\bglb\basel{\xi}{1},\, \ldots,\, \basel{\xi}{\mu}\bgrb \in \mygrp^{\mu}$ be plain message. For public key cryptography, the encrypted message is $\bglb\basel{\epsilon}{1},\, \ldots,\, \basel{\epsilon}{\mu}\bgrb  = P\bglb\basel{\xi}{1},\, \ldots,\, \basel{\xi}{\mu}\bgrb$, and the decryption is $P^{-1}\bglb\basel{\epsilon}{1},\, \ldots,\, \basel{\epsilon}{\mu}\bgrb$. For digital signature, the signed message is $\bglb\basel{\epsilon}{1},\, \ldots,\, \basel{\epsilon}{\mu}\bgrb  = P^{-1}\bglb\basel{\xi}{1},\, \ldots,\, \basel{\xi}{\mu}\bgrb$, and recovered message is $P\bglb\basel{\epsilon}{1},\, \ldots,\, \basel{\epsilon}{\mu}\bgrb$.

  In the remaining part of the section, it is assumed that $1 \leq \mu \leq \nu-1$.  Let $\kappa$ be the number of padding message elements in the hashing keys. It is assumed that the key generator ensures that a prospective owner of the pertinent keys is provided with an abundance of options for generating multivariate one-to-one mappings, whose inverse mappings are known only to the owner.  Let $\xx = (\basel{x}{1},\, \ldots,\, \basel{x}{\mu}) \in \mygrp^{\mu}$ be the plain message, $\yy = (\basel{y}{1},\, \ldots,\, \basel{y}{\nu})\in \eltSet^{\nu}$ be the encrypted or signed message, and  $\boldomega = (\basel{\omega}{1},\, \ldots,\, \basel{\omega}{\kappa}) \in \mygrp^{\kappa}$ be a padding message. For public key encryption, an injective mapping $P$ from $\mygrp^{\nu}$ into $\eltSet^{\nu}$ is chosen, while for digital signature, a surjective mapping $P$ from $\mygrp^{\nu}$ onto $\mygrp^{\mu}$ (in addition to two more surjective mappings) is chosen. Thus, for public key cryptography mapping, invertible parametric matrices in the most general form can be utilised, while for digital signature multivariate surjective or bijective mappings, only parametric permutation and diagonal matrices are employed.
  
   For public key cryptography, let $\lambda = \nu-\mu$, and let $P$ be an injective mapping from $\mygrp^{\nu}$ into $\eltSet^{\nu}$. Let $\yy \in \mygrp^{\nu}$ be the argument vector of the bijective mapping $P$. Then, the vector $\bglb \basel{f}{1}(\xx,\, \boldomega),\, \ldots, \basel{f}{\lambda}(\xx,\, \boldomega),\, \basel{x}{1},\, \ldots,\, \basel{x}{\mu}\bgrb$, for some hidden keys $\basel{f}{1}(\xx,\, \boldomega),\, \ldots, \basel{f}{\lambda}(\xx,\, \boldomega)$, is substituted for $\yy$ of the public key encryption mapping. Let  $F(\xx,\, \boldomega) = \bglb \basel{f}{1}(\xx,\, \boldomega),\, \ldots, \basel{f}{\lambda}(\xx,\, \boldomega)\bgrb$, and it is assumed that $F(\xx,\,\boldomega) \in \mygrp^{\lambda}$, for $\xx \in \mygrp^{\mu}$ and $\boldomega \in \mygrp^{\kappa}$. The information required to compute $P^{-1}(\boldvarepsilon)$, for $\boldvarepsilon \in \eltSet^{\nu}$, and the hidden hashing keys $F(\xx,\, \boldomega)$, for $\xx \in \mygrp^{\mu}$ and $\boldomega \in \mygrp^{\kappa}$, is saved in a private key back-substitution table $\BackSubstitutionTable$, while the mapping $P\bglb F(\xx,\, \boldomega),\, \xx\bgrb$ is saved in the public key lookup table $\LookupTable$. If the sender and receiver agree on $\boldomega$, and the encrypted message received is $\boldvarepsilon \in \mygrp^{\nu}$, then, with $(\zz,\, \xx) = P^{-1}(\boldvarepsilon)$, the receiver can ascertain data integrity by testing whether $F(\xx,\, \boldomega) = \zz$. It is possible to utilise $\boldomega$ as a session key in handshake protocols for repeated key negotiations.

  For digital signature, let $\kappa$ $\lambda$, $K$, $L$, $\mu$ and $\nu$ be positive integers, such that $K \leq \kappa$, $L \leq \lambda$ and $\nu \geq L+\mu$. Let $P$, $Q$ and $R$ be multivariate surjective mapping from $\mygrp^{\nu}$ onto $\mygrp^{L+\mu}$, from $\mygrp^{\kappa}$ onto $\mygrp^{K}$ and from $\mygrp^{\lambda}$ onto $\mygrp^{L}$, respectively.  The right inverse mappings of the stated multivariate surjective mappings are known only to the signer. Let $F(\xx,\, \boldomega) = \bglb \basel{f}{1}(\xx,\, \boldomega),\, \ldots,\, \basel{f}{L}(\xx,\,  \boldomega) \bgrb$. The components of the mapping $P$, corresponding to the plain message, are advertised in a public key signature verification table $\SignatureVerificationTable$, and the information for computing a right inverse of $P$ --- and, in general, all the information required by the signing algorithm --- is saved in a private key signature table $\SignatureTable$, for signing plain message. Now, for a plain message $\xx \in \mygrp^{\mu}$ and a padding message $\boldomega \in \mygrp^{\kappa}$, the signed message $\boldvarepsilon$ is obtained by applying  a right inverse mapping of $P$ on the instance $(\zz',\, \xx) \in \mygrp^{L+\mu}$, where $\zz' = F(\xx,\, \boldomega)$. The parameter $\zz \in \mygrp^{\lambda}$ is so chosen by the signer that $F(\xx,\,\boldomega) = R(\zz)$, by computing a right inverse of the multivariate surjective mapping $R$. For a plain message $\xx \in \mygrp^{\mu}$, the padding message $\boldomega \in \mygrp^{\kappa}$ is obtained by computing a right inverse, which is known only to the signer, of the multivariate surjective mapping $Q$ from $\mygrp^{\kappa}$ onto $\mygrp^{K}$, such that $Q(\boldomega) = \boldomega'$, where $\boldomega' \in \mygrp^{K}$ is agreed upon by the singer with a trusted authentication verifier TAV, for this particular signature transaction, as a first step in the signature generation procedure.  The plain message can be found by computing the components of the mapping $P$, that are advertised in the public key signature verification table $\SignatureVerificationTable$, for a signed message $\boldvarepsilon \in \mygrp^{\nu}$. For claiming the authenticity of the signature, the receiver of the signature needs to produce also $\zz$, which must be transmitted to the receiver by the signer. In addition to the signature verification table $\SignatureVerificationTable$, containing plain message components of the mapping $P$, another table $\SignatureAuthenticationTable$, called the signature authentication table, containing the full mapping $P$ and additional functions $H(\zz,\,\xx,\, \boldomega)$, with several components, {\em i.e.}, with values in $\eltSet^{\tau}$, for some positive integer $\tau$, is employed for signature authentication verification purpose, for which the signer meeds to provide $(\zz,\, \bolddelta,\, \boldomega)$, where $\bolddelta \in \eltSet^{\tau}$ is such that $\bolddelta = H\bglb \zz,\, \xx,\, \boldomega \bgrb$, at the signer end, and the signature authentication is verified by testing whether $H(\zz,\, \xx,\, \boldomega) = \bolddelta$ and $R(\zz) = F(\xx,\, \boldomega)$, by the verification authority, such that the vector $(\zz,\, \xx,\, \boldomega)$ satisfies additional conditions, such as $Q(\boldomega) = \boldomega'$, where $\boldomega'$ has been consented by the TAV for this signature.
  
  The signature authentication table $\SignatureAuthenticationTable$ is registered with a trusted authentication verifier (TAV), which is a public authority responsible for signature authentication verification purpose. The authentication information shared by the signer with TAV contains the multivariate mappings $P(\boldvarepsilon)$,  $F(\xx,\, \boldomega)$, $Q(\boldomega)$ and $R(\zz)$, where $P$, $Q$ and $R$ are surjective mappings from $\mygrp^{\nu}$ onto $\mygrp^{L+\mu}$, from $\mygrp^{\kappa}$ onto $\mygrp^{K}$ and from $\mygrp^{\lambda}$ onto $\mygrp^{L}$, respectively. The information required to compute any right inverse mappings of $P$, $Q$ and $R$ is known only to the owner of the signature keys, {\em i.e.}, the signer. The verification protocol at TAV side checks whether $Q(\boldomega)$ and $\xx$ meet certain obligations, and whether $R(\zz) = F(\xx,\, \boldomega)$, without knowing right inverse mappings of $Q$ and $R$. Now, for a particular plain message $\xx$ to be signed, the signer obtains an extra padding message $\boldomega' \in \mygrp^{K}$, with the consent of TAV, conforming to the predefined agreement for a valid padding message with TAV, and computes right inverse of $Q$ with $\boldomega'$ as the argument, to get the actual padding message $\boldomega \in \mygrp^{\kappa}$. Finally, with $\xx$ and $\boldomega$ having been chosen or computed, the signer generates $\zz \in \mygrp^{\lambda}$ by computing a right inverse of $R$ with $F(\xx,\, \boldomega)$ as the argument, and the signature itself by computing the inverse of the key mapping $P$, which is a multivariate surjective mapping from $ \mygrp^{\nu}$ into $\mygrp^{L+\mu}$. 
  
   It is possible to include $H(\zz,\, \xx,\, \boldomega)$ in the signature verification public key table $\SignatureVerificationTable$, in order to facilitate the receiver with a data integrity check, before approaching the TAV. This choice depends on the group of possible receivers and signers besides TAV. If the intended group of possible receivers is very large, such as external world, then it is convenient to reserve $H(\zz,\, \xx,\, \boldomega)$ to be present only in the signature authentication table $\SignatureAuthenticationTable$. In any case, the components of the map $P$ corresponding to $F(\xx,\, \boldomega)$ may be exclusively present only in the signature authentication table $\SignatureAuthenticationTable$, since disclosing this information to the public may lead to its speculation based on various observed values.
   
    In the proposed model of digital signature scheme, the signer approaches the TAV, with a request for generating a signed message for a specific purpose. The TAV issues consent for a particular extra padding message $\boldomega'$, for a period of validity along with a transaction number. The extra padding message $\boldomega'$ may contain a small gist of transaction details, encrypted by the key of TAV. Thus, the signer must request TAV, for the issuance of the extra padding message $\boldomega'$, by submitting a form containing a gist of transaction or signature details and its intended purpose. The TAV then generates an extra padding message  $\boldomega'$, transaction number and period of validity, and issues them to the signer. The signer is required to transmit the transaction number and period of validity to the intended receiver of the signature, who will have to produce these particulars to TAV for claiming the authenticity of the signed message. It may additionally be required that the claimants of the authenticity of a signed message will be required to furnish their signatures to TAV, with TAV and possibly also the sender bearing the role of the receiver, for a proof of the claim. 

  Multivariate surjective mappings can be realised as parametric mappings, which are bijective for some choice of parameter component values, and may be arbitrary mappings for some other choice of parameter values. The choice of parameters is known to the signer. For example, for the multivariate surjective mapping $Q$ from $\mygrp^{\kappa}$ onto $\mygrp^{K}$, $\kappa-K$ components of $\boldomega$ are taken in the argument vector of the partition of unity functions of section \ref{Sec-parametric-injective-mappings}, with $l =  \kappa-K$ and $m = K$, in the notation followed there. When combined with the partitioning methods of section \ref{sec-parametrization-of-permutation-polynomials}, for some partitions, with index $i$, the mappings $\basel{\zeta}{i}\bglb \basel{z}{1},\, \ldots,\, \basel{z}{l};\, \xx \bgrb$ are chosen to be bijective, and for the remaining, the mappings are arbitrary.

\section{\label{Sec-complexity-of-computing-nonparametric-inverses-of-parametric-multivariate-polynomial-mappings}Complexity Analysis of Computation of Inverse Mappings of Multivariate Mappings by Solving Simultaneous Multivariate Equations}
Model theory of fields and polynomial algebras
is extensively studied in mathematical logic 
[\cite{vanDalen:1994},  \cite{vandenDries:2000},   \cite{Hodges:1993}, 
 \cite{Marker:2000} and  \cite{MMP:1996}].
Let $\scalars$ be a field, and let {\small{$\ArithmeticExpressions(\scalars)$}}
be the set of arithmetic expressions without quantifiers,
obtained by collecting the expressions involving any number
of finitely many variables, constructed using parentheses and
the binary or unary arithmetic operators of addition $+$,
subtraction $-$, multiplication $\cdot$, possibly division $/$,
exponentiation $^k$, where $k$ is a positive integer, and
binary valued relational operator $=$ (and possibly other
relational operators such as $<$, \, $>$, \, $\leq$ and $\geq$).
The relational operators allow construction of
assertions that evaluate to anyone of the special symbolic constants
$\false$ and $\true$, represented by $0$ and $1$, respectively.
In the sequel, the variables assume values from $\scalars$,
the arithmetic expressions evaluate to values in $\scalars$,
as defined by the arithmetic operations in $\scalars$,
and the assertions evaluate to values in $\{0,\, 1\}$.
A variable taking values in $\{0,\, 1\}$ is a boolean
variable. The arithmetic expressions in
$\ArithmeticExpressions(\basel{\Integers}{2})$ 
are boolean expressions. For any field $\scalars$,
a boolean variable $x$ can be obtained from the equation
$x^{2}-x = 0$.  For boolean variables $x$ and $y$,
~$\lnot x$ can be represented by $1-x$, ~$x \wedge y$
by $x\cdot y$, ~$x \vee y$ by $1-(1-x)\cdot(1-y)$,
~$x\oplus y$ by $(x-y)^{2}$, ~$x \rightarrow y$ by
$1-x \cdot(1-y)$,  ~and $x \leftrightarrow y$ by $1-(x-y)^{2}$,
where $\lnot$ denotes the logical ``negation'',
$\wedge$ the logical  ``and'',
$\vee$ the logical ``or'',
$\oplus$ the logical ``exclusive or'',
$\rightarrow$ the logical ``implies'',
and $\leftrightarrow$ the logical ``implies and
is implied by''. The inequality operator, denoted by $\not =$,
is a secondary binary operator defined as 
the logical negation of the equality operator.
Let $ \QuantifiedArithmeticExpressions(\scalars)$
be the set of arithmetic expressions in which
some (none, some or all) variables are constrained
by ``existential'' $\exists$ or ``universal'' $\forall$
quantifiers. A variable constrained by a quantifier is
called a {\em bound} variable. A variable that is not
bound is called a {\em free} variable. An arithmetic
expression in which all the variables are free is a
quantifier free arithmetic expression, {\em i.e.},
an expression in $\ArithmeticExpressions(\scalars)$. 
A quantified arithmetic expression is in prenex normal
form, if all the quantifiers occur before the otherwise
quantifier free arithmetic expression, {\em i.e},
a quantified arithmetic expression of the form
{\small{$\forall \basel{y}{1}\, \ldots\, \forall \basel{y}{\basel{k}{1}} \,$
$ \exists \basel{x}{1}\,\ldots\,$
$ \forall \basel{y}{\basel{k}{i-1}+1}\, \ldots\, \forall \basel{y}{\basel{k}{i}} \, $
$\exists \basel{x}{i}\,\ldots\,$
$ \forall \basel{y}{\basel{k}{m-1}+1}\, \ldots\,$
$ \forall \basel{y}{\basel{k}{m}} \, $
$\exists \basel{x}{m}\,$
$\forall \basel{y}{\basel{k}{m}+1}\, \ldots\, \forall \basel{y}{n}$
~~$f(\basel{x}{1},\,\ldots,\, \basel{x}{m},\, 
\basel{y}{1},\, \ldots,\, \basel{y}{n})$},} where $m$ and $n$
are positive integers, and $\basel{k}{i}$, for $1 \leq i \leq m$,
are nonnegative integers such that $\basel{k}{i} \leq \basel{k}{i+1}$,
for $1 \leq i \leq m-1$, and $\basel{k}{m} \leq n$.
The variables $\basel{y}{j}$, $1 \leq j \leq n$, are 
{\em independent} variables, as they are bound to universal
quantifiers. The variable $\basel{x}{i}$ depends on the variables
$\basel{y}{j}$, $1 \leq j \leq \basel{k}{i}$, $1 \leq i \leq m$,
and  is a {\em dependent} bound variable. 
 A tuple $\bglb \basel {a}{1},\, \ldots,\, \basel{a}{i},\,
 \basel{b}{1},\, \ldots,\, \basel{b}{\basel{k}{i}}\bgrb
\in \scalars^{i+\basel{k}{i}}$, $1 \leq i \leq m$,
 is {\em feasible} to a quantified arithmetic expression
 in prenex normal form with no free variables as described before,
 if either $i = m$ and $f(\basel{a}{1},\,\ldots,\, \basel{a}{m},\, $
$\basel{b}{1},\, \ldots,\, \basel{b}{\basel{k}{m}},\,$
$\basel{y}{\basel{k}{m}+1},\, \ldots,\,\basel{y}{n})$
evaluates to $\true$, for $\basel{y}{\basel{k}{m}+1},\, \ldots,\,\basel{y}{n}$
$ \in $
$\scalars$,
or $1 \leq i \leq m-1$ and
for $\basel{y}{\basel{k}{i}+1},\, \ldots,\,\basel{y}{\basel{k}{i+1}} \in \scalars$,
and for some $\basel{x}{i+1} \in \scalars$, that may depend on 
$ \basel {a}{1},\, \ldots,\, \basel{a}{i},\,$,
$\basel{b}{1},\, \ldots,\, \basel{b}{\basel{k}{i}},\, $
$\basel{y}{\basel{k}{i}+1},\, \ldots,\,\basel{y}{\basel{k}{i+1}}$
$ \in \scalars$, 
 each tuple
{\small{$(\basel{a}{1},\, \ldots,\, \basel{a}{i},\, \basel{x}{i+1},\,$
$ \basel{b}{1},\, \ldots,\, \basel{b}{\basel{k}{i}}, \, $
$ \basel{y}{\basel{k}{i}+1},\, \ldots,\, \basel{y}{\basel{k}{i+1}})$}}
is feasible. If for every $\basel{b}{1},\, \ldots,\, \basel{b}{\basel{k}{1}}$
$ \in $
$\scalars$, there exists $\basel{a}{1} \in \scalars$,
 such that the tuple $\bglb \basel {a}{1},\, 
  \basel{b}{1},\, \ldots,\, \basel{b}{\basel{k}{1}}\bgrb$
 is feasible, then the given instance of binary valued
 quantified arithmetic expression is {\em satisfiable}.
 The evaluation problem for quantified boolean expressions
 in prenex normal form with no free variables in 
 $\QuantifiedArithmeticExpressions(\scalars)$ is
 to find whether the given input instance is satisfiable.
 Let $\SatisfiableQuantifiedArithmeticExpressions(\scalars)
 \subseteq \QuantifiedArithmeticExpressions(\scalars)$ be the set
  of satisfiable binary valued quantified arithmetic expressions
 ({\em i.e.}, quantified arithmetic assertions) in prenex normal form
 with no free variables that evaluate to $\true$. Let
 $\QBF$ and $\QSAT$ be $\QuantifiedArithmeticExpressions(\basel{Z}{2})$
 and $\SatisfiableQuantifiedArithmeticExpressions(\basel{Z}{2})$,
 respectively. By the previous discussion, every boolean expression in $\QBF$,
 analogously in $\QSAT$, can be represented by some arithmetic expression
 in $\QuantifiedArithmeticExpressions(\scalars)$, analogously in
$\SatisfiableQuantifiedArithmeticExpressions(\scalars)$,
with equality binary relation, for any field $\scalars$.
The evaluation problem for quantified boolean expressions
in prenex normal form with no free variables in $\QBF$ is
$\PSPACE$-complete, where $\PSPACE$ is the set of formal
languages acceptable in polynomial space [\cite{HMU:2007}].

It may be recalled that, by convention, the binary value of $\true$
is taken to be $1$ and that of $\false$ is $0$.  It is occasionally 
convenient to interpret $\true$ to be ``nonzero'' and $\false$ to be
the value $0$.

\subsection{\label{Sec-CSP}Constraint Satisfaction and Quantifier Elimination Problems}
Let 
\hfill{\small{$\forall \basel{y}{1}\, \ldots\, \forall \basel{y}{\basel{k}{1}} 
\, \exists \basel{x}{1}\,\ldots\,
 \forall \basel{y}{\basel{k}{i-1}+1}\, \ldots\, \forall \basel{y}{\basel{k}{i}} 
\, \exists \basel{x}{i}\,\ldots\,
 \forall \basel{y}{\basel{k}{m-1}+1}\, \ldots\, \forall \basel{y}{\basel{k}{m}} 
\, \exists \basel{x}{m}$}}
{\small{$\, \forall \basel{y}{\basel{k}{m}+1}\, \ldots\, \forall \basel{y}{n}
~~f(\basel{x}{1},\,\ldots,\, \basel{x}{m},\, 
\basel{y}{1},\, \ldots,\, \basel{y}{n})$}} be an instance
in {\small{$\QuantifiedArithmeticExpressions(\scalars)$},}
where $m$ and $n$ are positive integers, and $\basel{k}{i}$,
for $1 \leq i \leq m$, are nonnegative integers such that
$\basel{k}{i} \leq \basel{k}{i+1}$, for $1 \leq i \leq m-1$,
and $\basel{k}{m} \leq n$. A functional solution for the given 
instance of constraint satisfaction problem is a sequence of
quantifier free arithmetic expressions  
$\basel{g}{1}(\basel{y}{1},\,\ldots,\,\basel{y}{\basel{k}{1}})$
and $\basel{g}{i} (\basel{x}{1},\, \ldots, \basel{x}{i-1}, \,$
$\basel{y}{1},\,\ldots,\,\basel{y}{\basel{k}{i}})$, $2 \leq i \leq m$,
in $\ArithmeticExpressions \bglb \scalars \bgrb$, such that
$\forall \basel{y}{1}\, \ldots\, \forall \basel{y}{n}~$
$f(\basel{x}{1},\,\ldots,\, \basel{x}{m},\, $
$\basel{y}{1},\, \ldots,\, \basel{y}{n}) $
$ ~ = ~ \true$, where
{\small{$\basel{x}{1} = \basel{g}{1}(\basel{y}{1},\,\ldots,\,\basel{y}{\basel{k}{1}})$}}
and 
{\small{$ \basel{x}{i} = \basel{g}{i} (\basel{x}{1},\, \ldots, \basel{x}{i-1}, \,$
$ \basel{y}{1},\,\ldots,\,\basel{y}{\basel{k}{i}})$, $2 \leq i \leq m$}.}
 The constraint satisfaction problem is feasible, if it has a functional solution
in quantifier free arithmetic expressions.

Let $\powerset(\scalars)$ be a set of parametric subsets of $\scalars$,
parametrised by variables assuming values in $\scalars$, such that the
binary valued characteristic functions of the sets are assertions in
$\ArithmeticExpressions(\scalars)$. For an instance in
$\QuantifiedArithmeticExpressions(\scalars)$,
the quantifier elimination problem is to compute parametric sets
{\small{$\basel{G}{i}(\basel{x}{1},\,$
$ \ldots, \,$
$ \basel{x}{i-1},\,$ 
 $\basel{y}{1},\,\ldots,\, \basel{y}{\basel{k}{i}})$
in $\powerset(\scalars)$},} 
for {\small{$\basel{x}{1},\, \ldots,\,$
$ \basel{x}{i-1},\,$
$\basel{y}{1},\,\ldots,\,$
$ \basel{y}{\basel{k}{i}}$
$ \in $
$\scalars$},}  such that
\begin{small}
\begin{eqnarray*}
&& \ltab \ltab \basel{G}{i}(\basel{x}{1},\, \ldots, \, \basel{x}{i-1},\, 
 \basel{y}{1},\,\ldots,\, \basel{y}{\basel{k}{i}})
~~~~ = \\
&&  \bglc \basel{x}{i} \in \scalars\,:\, 
 (\basel{x}{1},\, \ldots, \, \basel{x}{i},\, 
 \basel{y}{1},\,\ldots,\, \basel{y}{\basel{k}{i}}) ~
\textrm{is feasible to the given instance} \bgrc 
\end{eqnarray*}
\end{small}
\lspace for $1 \leq i \leq m$. Now, for an instance in
$\QuantifiedArithmeticExpressions(\scalars)$,
the problem of computing feasible parameter sets is to compute, 
for {\small{$\basel{x}{1}, \ldots,\, \basel{x}{i-1},\,$
$\basel{y}{1},\,\ldots,\, \basel{y}{\basel{k}{i}}$
$ \in $
$\scalars$}}, the characteristic (indicator) functions
{\small{$\basel{\theta}{i}(\basel{x}{1},\, \ldots, \, \basel{x}{i-1},\, $
$ \basel{y}{1},\,\ldots,\, \basel{y}{\basel{k}{i}})$
in $\ArithmeticExpressions(\scalars)$},}
such that the function $\basel{\theta}{i}(\basel{x}{1},\, \ldots, \, \basel{x}{i-1},\, $
$ \basel{y}{1},\,\ldots,\, \basel{y}{\basel{k}{i}})$
 evaluates to nonzero, if and only if the corresponding set
$\basel{G}{i}(\basel{x}{1},\, \ldots, \, \basel{x}{i-1},\, $
$ \basel{y}{1},\,\ldots,\, \basel{y}{\basel{k}{i}})$
is nonempty, for $1 \leq i \leq m$.
Set solutions can be enumerated by backtracking method [\cite{HSR:2007}].
In the definition of $\ArithmeticExpressions(\scalars)$,
relational assertions may also be present.

\begin{theorem}
\label{hardness-of-prenex-normal-form-feasibility-set-problem-for-QBF}
The problem of computing feasible parameter sets for instances in prenex normal form in $\QuantifiedArithmeticExpressions (\scalars)$ is $\PSPACE$-hard.
\end{theorem}
\proof
Let 
\hfill{\small{$\forall \basel{y}{1}\, \ldots\, \forall \basel{y}{\basel{k}{1}} 
\, \exists \basel{x}{1}\,\ldots\,
 \forall \basel{y}{\basel{k}{i-1}+1}\, \ldots\, \forall \basel{y}{\basel{k}{i}} 
\, \exists \basel{x}{i}\,\ldots\,
 \forall \basel{y}{\basel{k}{m-1}+1}\, \ldots\, \forall \basel{y}{\basel{k}{m}} 
\, \exists \basel{x}{m}$}}
{\small{$\, \forall \basel{y}{\basel{k}{m}+1}\, \ldots\, \forall \basel{y}{n}
~~f(\basel{x}{1},\,\ldots,\, \basel{x}{m},\, 
\basel{y}{1},\, \ldots,\, \basel{y}{n})$}} be an instance
in $\QuantifiedArithmeticExpressions(\scalars)$,
where $m$ and $n$ are positive integers, and $\basel{k}{i}$,
for $1 \leq i \leq m$, are nonnegative integers such that
$\basel{k}{i} \leq \basel{k}{i+1}$, for $1 \leq i \leq m-1$,
and $\basel{k}{m} \leq n$.
Then, for the instance 
 \hfill{\small{$\exists \basel{x}{0}\, \forall \basel{y}{1}\, \ldots\, \forall \basel{y}{\basel{k}{1}} 
\, \exists \basel{x}{1}\,$
$\ldots\,$
$ \forall \basel{y}{\basel{k}{i-1}+1}\, \ldots\, \forall \basel{y}{\basel{k}{i}} 
\, \exists \basel{x}{i}\,$
$\ldots\,$
$ \forall \basel{y}{\basel{k}{m-1}+1}\, \ldots\, \forall \basel{y}{\basel{k}{m}} 
\, \exists \basel{x}{m} \,$
$ \forall \basel{y}{\basel{k}{m}+1}\, \ldots\, \forall \basel{y}{n}$
$~~\bglb ~ \basel{x}{0} $
$\cdot $
$ f(\basel{x}{1},\,\ldots,\, \basel{x}{m},\, $
$\basel{y}{1},\, \ldots,\, \basel{y}{n}) ~ \bgrb$}} 
in $\QuantifiedArithmeticExpressions(\scalars)$,
the characteristic function $\basel{\chi}{0}$ 
for the feasible parameter set of $\basel{x}{0}$ is nonzero 
if and only if the originally given instance is feasible. \qed

\subsection{\label{Sec-simultaneous-multivariate-polynomial-equations}Simultaneous Multivariate Polynomial Equations}

Let $\powerset(\scalars)$ be a set of parametric subsets of $\scalars$,
parametrised by variables assuming values in $\scalars$, such that the
binary valued characteristic functions of the sets are assertions in
$\ArithmeticExpressions(\scalars)$. Let $l,\, m, \, n \in \PositiveIntegers$
and $\basel{f}{t}\bglb \basel{x}{1},\,\ldots,\, \basel{x}{m}, \,$
$ \basel{y}{1},\, \ldots, \, \basel{y}{n}\bgrb $
$ \in $
$\ArithmeticExpressions(\scalars)$, for $1 \leq t \leq l$, be arithmetic expressions. A system of (multivariate) polynomial equations is the following:
\begin{equation}
\basel{f}{t}\bglb \basel{x}{1},\,\ldots,\, \basel{x}{m},\, \basel{y}{1},\, \ldots, \, \basel{y}{n}\bgrb  \tab = \tab 0\,,
  \tab\tab 1 \leq t \leq l\,, \label{System-of-polynomial-equations}
\end{equation}
where $\basel{y}{j}$, $1 \leq j \leq n$, are independent variables,
and $\basel{x}{i}$, $1 \leq i \leq m$, are dependent variables,
assuming values from $\scalars$, both specified as part of an instance. 
The expressions in (\ref{System-of-polynomial-equations}) may also involve relational assertions.

A tuple $\bglb \basel{a}{1},\, \ldots, \, \basel{a}{i}, \, $
$ \basel{b}{1},\, \ldots, \, \basel{b}{n}\bgrb$ is feasible to (\ref{System-of-polynomial-equations}), if either (1) $i = m$ and 
(\ref{System-of-polynomial-equations}) holds with $\basel{x}{r} = \basel{a}{r}$, for $1 \leq r \leq m$, and $\basel{y}{j} = \basel{b}{j}$, for $1 \leq j \leq n$, or (2) $1 \leq i \leq m-1$, and 
 $\bglb \basel{a}{1},\, \ldots, \, \basel{a}{i}, \, \basel{a}{i+1},\,$
 $  \basel{b}{1},\, \ldots, \, \basel{b}{n}\bgrb$ is feasible for some $\basel{a}{i+1}$,
 possibly depending on $\bglb \basel{a}{1},\, \ldots, \, \basel{a}{i},  \, \basel{b}{1},\, \ldots, \, \basel{b}{n}\bgrb$. Let $\powerset\bglb \scalars \bgrb$ be the collection of admissible subsets of $\scalars$, whose indicator functions are in $\ArithmeticExpressions \bglb \scalars \bgrb$. A complete solution to (\ref{System-of-polynomial-equations}) are parametric sets
{\small{$\basel{H}{i}\bglb\basel{a}{1},\, \ldots,\, \basel{a}{i-1},\,$
$\basel{y}{1},\, \ldots,\, \basel{y}{n}\bgrb$
$ \in $
$\powerset\bglb \scalars \bgrb$},} such that 
{\small{$\basel{H}{i}\bglb\basel{a}{1},\, \ldots,\, \basel{a}{i-1},\,$
$\basel{y}{1},\, \ldots,\, \basel{y}{n}\bgrb$
$ = $
$\bglc \basel{a}{i} \in \scalars\, :\, $
$\bglb\basel{a}{1},\, \ldots,\, \basel{a}{i-1},\, \basel{a}{i},\, $
$\basel{y}{1},\, \ldots,\, \basel{y}{n}\bgrb $
$~~ \textrm{is feasible} \bgrc$},} for  $1 \leq i \leq m$. 
Now, the problem of computing feasible parameter sets
 to (\ref{System-of-polynomial-equations}) is to compute
the quantifier free characteristic (indicator) functions 
{\small{$\basel{\vartheta}{i}\bglb\basel{x}{1},\, \ldots,\, \basel{x}{i-1}, \,$
$ \basel{y}{1},\, \ldots, \, \basel{y}{n}\bgrb$}}
in $\ArithmeticExpressions(\scalars)$, such that 
{\small{$\basel{\vartheta}{i}\bglb\basel{x}{1},\, \ldots,\, \basel{x}{i-1}, \,$
$ \basel{y}{1},\, \ldots, \, \basel{y}{n}\bgrb$
$ \neq $
$0$}}
if and only if 
   {\small{$\basel{H}{i}\bglb\basel{x}{1},\, \ldots,\, \basel{x}{i-1},\,$
$\basel{y}{1},\, \ldots,\, \basel{y}{n}\bgrb \neq \emptyset $},}
for $1 \leq i \leq m$.

In the system of multivariate equations of (\ref{System-of-polynomial-equations}), the ordering of the variables $\basel{x}{1},\ldots, \basel{x}{m}$ appears specified. However, this ordering can be made innocuous by additional constraints as follows:
\begin{small}
\begin{eqnarray*}
&& \tab \tab \basel{v^{2}}{i,\, j} ~~ = ~~ \basel{v}{i,\,j} \tab \textrm{and} \tab \basel{w^{2}}{i,\, j} ~~ = ~~ \basel{w}{i,\,j}\,, \tab 1 \leq i,\, j \leq m\\
&& \sum_{j = 1}^{m} \basel{v}{i,\,j} ~~ = ~~ 1 \tab \textrm{and} \tab
\basel{v}{i,\, j} \cdot \basel{v}{i,\, k} ~~ = ~~ 0\, , \tab 1 \leq i,\, j,\, k \leq m ~~ \textrm{and} ~~ j \neq k\\
&& \sum_{i = 1}^{m} \basel{v}{i,\,j} ~~ = ~~ 1 \tab \textrm{and} \tab 
\basel{v}{i,\, j} \cdot \basel{v}{k,\, j} ~~ = ~~ 0\, , \tab 1 \leq i,\, j,\, k \leq m ~~ \textrm{and} ~~ i \neq k\\
&&\\
&& \sum_{j = 1}^{m} \basel{w}{i,\,j} ~~ = ~~ 1 \tab \textrm{and} \tab
\basel{w}{i,\, j} \cdot \basel{w}{i,\, k} ~~ = ~~ 0\, , \tab 1 \leq i,\, j,\, k \leq m ~~ \textrm{and} ~~ j \neq k\\
&& \sum_{i = 1}^{m} \basel{w}{i,\,j} ~~ = ~~ 1 \tab \textrm{and} \tab 
\basel{w}{i,\, j} \cdot \basel{w}{k,\, j} ~~ = ~~ 0\, , \tab 1 \leq i,\, j,\, k \leq m ~~ \textrm{and} ~~ i \neq k\\
&&\\
&&\ltab \ltab ~~   \left[ \begin{array}{cccc}
\basel{v}{1,\,1}& \basel{v}{1,\,2}& \ldots & \basel{v}{1,\,m}\\
\basel{v}{2,\,1}& \basel{v}{2,\,2}& \ldots & \basel{v}{2,\,m}\\
\vdots & \vdots & \vdots  & \vdots \\
\basel{v}{m,\,1}& \basel{v}{m,\,2}& \ldots & \basel{v}{m,\,m}\\
\end{array} \right] 
\left[
\begin{array}{c}
\basel{x}{1}\\
\basel{x}{2}\\
\vdots\\
\basel{x}{m}\\
\end{array}\right]   = 
\left[ \begin{array}{cccc}
\basel{w}{1,\,1}& \basel{w}{1,\,2}& \ldots & \basel{w}{1,\,m}\\
\basel{w}{2,\,1}& \basel{w}{2,\,2}& \ldots & \basel{w}{2,\,m}\\
\vdots & \vdots & \vdots  & \vdots \\
\basel{w}{m,\,1}& \basel{w}{m,\,2}& \ldots & \basel{w}{m,\,m}\\
\end{array} \right]  
\left[
\begin{array}{c}
\basel{x}{m+1}\\
\basel{x}{m+2}\\
\vdots\\
\basel{x}{2m}\\
\end{array}\right]   \\
&&\\
&&  \ltab ~~ \mathrm{and}  \tab \tab  
\basel{f}{i}\bglb \basel{x}{m+1},\,\ldots,\, \basel{x}{2m}, \, \basel{y}{1},\, \ldots, \, \basel{y}{n}\bgrb  ~~~~ = ~~~~ 0\,,
  \tab\tab 1 \leq i \leq l
\end{eqnarray*}
\end{small}
\lspace where $\basel{y}{j}$, $1 \leq j \leq n$, are independent variables, and all the remaining variables are dependent variables. The ordering is concealed by allowing the system to choose an appropriate ordering of the variables $\basel{x}{m+1},\, \ldots, \,\basel{x}{2m}$, while allowing $\basel{x}{1}, \ldots,\, \basel{x}{m}$ to appear in the specified order. In the above set of constraints, for  each row of the matrix 
$\basel{\left[\basel{v}{i,\,j}\right]}{1 \leq i,\, j \leq m}$, for the constraints on $i$, $1 \leq i \leq m$, and for each column of the matrix $\basel{\left[\basel{v}{i,\,j}\right]}{1 \leq i,\, j \leq m}$, for the constraints on $j$, $1 \leq j \leq m$, the first constraint requires at least one entry of $1$,
and the second constraint requires $(m-1)$ entries of $0$, in the respective row or column,
and the matrix $\basel{\left[\basel{v}{i,\,j}\right]}{1 \leq i,\, j \leq m}$ is a permutation matrix. Similarly, the matrix 
$\basel{\left[\basel{w}{i,\,j}\right]}{1 \leq i,\, j \leq m}$
is also a permutation matrix.

 For indexes $i$ and $j$, where $1 \leq i < j \leq l$,
an inequation condition of the form
$\basel{f}{i}\bglb \basel{x}{1},\,\ldots,\, \basel{x}{m},\,
 \basel{y}{1},\, \ldots, \, \basel{y}{n}\bgrb  ~ \neq ~0$
  can be converted into an equation by the addition of a 
new dependent variable $\basel{z}{i}$ and the condition
$\bglb \basel{z}{i} \cdot \basel{f}{i}( \basel{x}{1},\,\ldots,\, \basel{x}{m},\,$
$ \basel{y}{1},\, \ldots, \, \basel{y}{n} )\bgrb  -1 $
$~ = ~ 0$. 
 After converting inequations into equations, any newly introduced variables are
assigned precedence ordering, that is usually subsequent to the dependent variables in the
original system, or may also be left unspecified.
The disjunction of equations 
$\basel{f}{i}( \basel{x}{1},\,\ldots,\, \basel{x}{m},\,$
$ \basel{y}{1},\, \ldots, \, \basel{y}{n}) $
$ = 0$ or
$\basel{f}{j} ( \basel{x}{1},\,\ldots,\, \basel{x}{m},\, $
$\basel{y}{1},\, \ldots, \, \basel{y}{n}) $
$  = 0$
can be replaced with 
$\bglb  \basel{f}{i}(\basel{x}{1},\,\ldots,\, \basel{x}{m},\,$
$ \basel{y}{1},\, \ldots, \, \basel{y}{n}) $
$ \cdot $
$\basel{f}{j} ( \basel{x}{1},\,\ldots,\, \basel{x}{m},\, $
$\basel{y}{1},\, \ldots, \, \basel{y}{n})\bgrb $
$  = 0$, to turn the disjunction into simultaneity.

The problems of computing solutions, complete solutions and 
characteristic (indicator) functions of feasible parameter sets
for selective dependent variables, $\basel{x}{\basel{i}{j}}$,
for $1 \leq j \leq r \leq m$, can be easily defined,
where $1 \leq \basel{i}{j} < \basel{i}{j+1}$,
for $1 \leq j \leq r-1$ and  $1 \leq r \leq m$,
requiring the correspondingly stated solutions only
for these dependent variables, treating the remaining
dependent variables as bound by existential quantifiers.
A procedure for solving a system of simultaneous multivariate
equations may solve for all the dependent variables, while 
producing output solutions only for the selective variables.
The mentioning of existential quantifiers may be bypassed,
for the solutions of selective variables. 

\begin{theorem}
\label{polynomial-time-reducibility-of-QBF-to-SMPE}
The combined problem of computing feasible parameter sets and
quantifier elimination for the instances in
$\QuantifiedArithmeticExpressions(\scalars)$, that are with
no free variables and in prenex normal form, is polynomial time
subroutine reducible to that of simultaneous multivariate equations
for selective variables over $\scalars$.
\end{theorem}
\proof
 Let 
\hfill{\small{$\forall \basel{y}{1}\, \ldots\, \forall \basel{y}{\basel{k}{1}} 
\, \exists \basel{x}{1}\,\ldots\,
 \forall \basel{y}{\basel{k}{i-1}+1}\, \ldots\, \forall \basel{y}{\basel{k}{i}} 
\, \exists \basel{x}{i}\,\ldots\,
 \forall \basel{y}{\basel{k}{m-1}+1}\, \ldots\, \forall \basel{y}{\basel{k}{m}} 
\, \exists \basel{x}{m}$}}
{\small{$\, \forall \basel{y}{\basel{k}{m}+1}\, \ldots\, \forall \basel{y}{n}
~~f(\basel{x}{1},\,\ldots,\, \basel{x}{m},\, 
\basel{y}{1},\, \ldots,\, \basel{y}{n})$}} be an
instance in  $\QuantifiedArithmeticExpressions(\scalars)$
in prenex normal form with no free variables, 
where $m$ and $n$ are positive integers, and $\basel{k}{i}$,
for $1 \leq i \leq m$, are nonnegative integers such that
$\basel{k}{i} \leq \basel{k}{i+1}$, for $1 \leq i \leq m-1$,
and $\basel{k}{m} \leq n$.

Let $\basel{\chi}{i}(\basel{x}{1},\,\ldots,\, \basel{x}{i},\, 
\basel{y}{1},\, \ldots,\, \basel{y}{\basel{k}{i}})$ 
be the characteristic function of the feasible parameter set
for the selective variable $\basel{z}{i,\,\basel{k}{i}+1}$,
whenever $\basel{k}{i} < \basel{k}{i+1}$, 
for the instance of simultaneous multivariate equations
\begin{small}
\[
\basel{\phi}{i+1}(\basel{x}{1},\,\ldots,\, \basel{x}{i},\, 
\basel{y}{1},\, \ldots,\, \basel{y}{\basel{k}{i}},\, \basel{z}{i,\,\basel{k}{i}+1},\, \ldots,\, \basel{z}{i,\, \basel{k}{i+1}}) ~~ = ~~ 0
\]
\end{small}
\lspace with {\small{$\basel{x}{1},\,\ldots,\, \basel{x}{i},\, ~
\basel{y}{1},\, \ldots,\, \basel{y}{\basel{k}{i}}$}}
as the independent variables and
{\small{$ \basel{z}{i,\,\basel{k}{i}+1},\,$
$ \ldots,\,$
$ \basel{z}{i,\, \basel{k}{i+1}}$}}
as the dependent variables with the specified mandatory ordering,
for $i = m, \, \ldots,\, 1$, in the descending order, where $\basel{k}{m+1} = n$ and
\begin{small}
\begin{eqnarray*}
&& \basel{\phi}{m+1}(\basel{x}{1},\,\ldots,\, \basel{x}{m},\, 
\basel{y}{1},\, \ldots,\, \basel{y}{\basel{k}{m}},\, \basel{z}{m,\,\basel{k}{m}+1},\, \ldots,\, \basel{z}{m,\, \basel{k}{m+1}}) ~~ = \\
&& \tab \tab  f(\basel{x}{1},\,\ldots,\, \basel{x}{m},\, 
\basel{y}{1},\, \ldots,\, \basel{y}{\basel{k}{m}},\, \basel{z}{m,\,\basel{k}{m}+1},\, \ldots,\, \basel{z}{m,\, \basel{k}{m+1}})
\end{eqnarray*}
\end{small}
\lspace If {\small{$\basel{k}{i} = \basel{k}{i+1}$},} then let 
{\small{$\basel{\chi}{i}(\basel{x}{1},\,\ldots,\, \basel{x}{i},\, $
$\basel{y}{1},\, \ldots,\, \basel{y}{\basel{k}{i}})$}} 
be {\small{$\lnot \basel{\phi}{i+1}(\basel{x}{1},\,\ldots,\, \basel{x}{i},\,$ 
$\basel{y}{1},\, \ldots,\, \basel{y}{\basel{k}{i}})$},}
for $1 \leq i \leq m$.

After obtaining 
$ \basel{\chi}{i}(\basel{x}{1},\,\ldots,\, \basel{x}{i},\, 
\basel{y}{1},\, \ldots,\, \basel{y}{\basel{k}{i}})$, for some
index $i$, where $1 \leq i \leq m$ --- such that the condition 
$\basel{\chi}{i}(\basel{x}{1},\,\ldots,\, \basel{x}{i},\, $
$\basel{y}{1},\, \ldots,\, \basel{y}{\basel{k}{i}}) = 0$ 
is the defining relation of the solution set
$\basel{G}{i}(\basel{x}{1},\,\ldots,\, \basel{x}{i-1},\, $
$\basel{y}{1},\, \ldots,\, \basel{y}{\basel{k}{i}})$,
for the variable $\basel{x}{i}$, with 
$\basel{x}{1},\,\ldots,\, \basel{x}{i-1},\, $
$\basel{y}{1},\, \ldots,\, \basel{y}{\basel{k}{i}}$
as the parameters, for the quantifier elimination problem,
as will be shown in a subsequent paragraph ---
let $ \basel{\phi}{i}(\basel{x}{1},\,\ldots,\, \basel{x}{i-1},\, $
$\basel{y}{1},\, \ldots,\, \basel{y}{\basel{k}{i}})$ be
the characteristic function of the feasible parameter set of the
variable $\basel{x}{i}$, for the instance of simultaneous
multivariate equations
\begin{small}
\[
 \basel{\chi}{i}(\basel{x}{1},\,\ldots,\, \basel{x}{i},\, 
\basel{y}{1},\, \ldots,\, \basel{y}{\basel{k}{i}}) ~~ = ~~ 0
\]
\end{small}
\lspace with $\basel{x}{1},\,\ldots,\, \basel{x}{i-1},\, $
$\basel{y}{1},\, \ldots,\, \basel{y}{\basel{k}{i}}$
as the independent variables and $\basel{x}{i}$ as
the dependent variable, for $i = m,\, \ldots,\, 1$,
in the descending order.

  It may be observed that the condition 
{\small{$\basel{\chi}{m}(\basel{x}{1},\,\ldots,\, \basel{x}{m},\, $
$\basel{y}{1},\, \ldots,\, \basel{y}{\basel{k}{m}}) = 0$}} 
is the defining relation of the solution set
{\small{$\basel{G}{m}(\basel{x}{1},\,\ldots,\, \basel{x}{m-1},\, $
$\basel{y}{1},\, \ldots,\, \basel{y}{\basel{k}{m}})$},}
for the variable $\basel{x}{m}$, with 
{\small{$\basel{x}{1},\,\ldots,\, \basel{x}{m-1},\, $
$\basel{y}{1},\, \ldots,\, \basel{y}{\basel{k}{m}}$}}
as the parameters, for the quantifier elimination problem,
and that the condition 
{\small{$\basel{\phi}{i}(\basel{x}{1},\,\ldots,\, \basel{x}{i-1},\, $
$\basel{y}{1},\, \ldots,\, \basel{y}{\basel{k}{i}}) \neq 0$}} 
is the defining relation for some parametric set of values
of {\small{$\basel{x}{i-1}$},} denoted by 
{\small{$\basel{H}{i-1}(\basel{x}{1},\,\ldots,\, \basel{x}{i-2},\, $
$\basel{y}{1},\, \ldots,\, \basel{y}{\basel{k}{i}})$},}   
to be nonempty, such that the set 
{\small{$\basel{G}{i-1}(\basel{x}{1},\,\ldots,\, \basel{x}{i-2},\, $
$\basel{y}{1},\, \ldots,\, \basel{y}{\basel{k}{i-1}})$}}
is given by 
{\small{$\bigcap_{\basel{y}{\basel{k}{i-1}+1}} $
$\cdots$
$ \bigcap_{\basel{y}{\basel{k}{i}}}$
$ \basel{H}{i-1}(\basel{x}{1},\,$
$\ldots,\,$
$ \basel{x}{i-2},\,$ 
$\basel{y}{1},\,$
$ \ldots,\,$
$ \basel{y}{\basel{k}{i}})$},} 
 for  {\small{$ \basel{x}{1},\,$
 $\ldots,\, $
 $\basel{x}{i-2},\,$ 
$\basel{y}{1},\,$
$ \ldots,\, $
$\basel{y}{\basel{k}{i}}$
$ \in $
$\scalars$}} and {\small{$2 \leq i \leq m+1$}.} 

 Now, the function  
{\small{ $\basel{\chi}{i-1}(\basel{x}{1},\,\ldots,\, \basel{x}{i-1},\, $
$\basel{y}{1},\, \ldots,\, \basel{y}{\basel{k}{i-1}})$},} for {\small{$2 \leq i \leq m+1$},}
is the characteristic function of the feasible parameter set
for the selective variable {\small{$\basel{z}{i-1,\,\basel{k}{i-1}+1}$}}
for the instance of simultaneous multivariate equations
\begin{small}
\[
\basel{\phi}{i}(\basel{x}{1},\,\ldots,\, \basel{x}{i-1},\, 
\basel{y}{1},\, \ldots,\, \basel{y}{\basel{k}{i-1}},\, \basel{z}{i-1,\,\basel{k}{i-1}+1},\, \ldots,\, \basel{z}{i-1,\, \basel{k}{i}}) ~~ = ~~ 0
\]
\end{small}
\lspace with {\small{$\basel{x}{1},\, \ldots,\,\basel{x}{i-1},\, 
\basel{y}{1},\, \ldots,\, \basel{y}{\basel{k}{i-1}}$}}
as the independent variables and
{\small{$ \basel{z}{i-1, \,\basel{k}{i-1}+1},\,$
$\ldots,\, $
$\basel{z}{i-1,\, \basel{k}{i}}$}}
as the dependent variables with the specified mandatory ordering.
Thus, the condition 
{\small{ $\basel{\chi}{i-1}(\basel{x}{1},\,\ldots,\, \basel{x}{i-1},\, 
\basel{y}{1},\, \ldots,\, \basel{y}{\basel{k}{i-1}})  =  0$}}
is the defining relation for the solution set, 
namely, {\small{$\basel{G}{i-1}(\basel{x}{1},\,\ldots,\, \basel{x}{i-2},\, 
\basel{y}{1},\, \ldots,\, \basel{y}{\basel{k}{i-1}})$},}
 for the variable {\small{$\basel{x}{i-1}$}}
in the given instance of quantifier elimination problem,
 with 
{\small{$\basel{x}{1},\,\ldots,\, \basel{x}{i-2}, \, $
$\basel{y}{1},\, \ldots,\, \basel{y}{\basel{k}{i-1}}$}}
as the parameters. \qed

\section{\label{Sec-Security-Analysis}Security Analysis}

   The classical analysis of multivariate simultaneous equations can be applied only to polynomial equations [\cite{vanDalen:1994},  \cite{vandenDries:2000},  \cite{Manin:2010},  \cite{Marker:2000},  \cite{MMP:1996} and \cite{Tarski:1951}], and the Gr\"{o}bner basis analysis [\cite{Buchberger:1965},  \cite{Faugere:1999} and  \cite{Faugere:2002}] is employed as the main practical tool. A general purpose method for solving multivariate mappings involving functions as exponents is not known as yet. For a security that is immune to threats from Gr\"{o}bner basis analysis, parametric injective mappings from $\mygrp^{\mu}$ into $\eltSet^{\nu}$, with $\kappa$ parameters, for $\mygrp = \nonzeroscalars$, $\eltSet = \scalars$ and $\mu,\, \nu,\, \kappa \in \PositiveIntegers$, where $1 \leq \mu \leq \nu$ and $\scalars$ is a finite field, with component mappings taken as expressions from {\small{$\multiexpressions{\scalars}{x}{\omega}{\mu}{\kappa}$},} restricting values of $\basel{x}{i}$ and $\basel{\omega}{j}$ to $\nonzeroscalars$, for $1 \leq i \leq \mu$ and $1 \leq j \leq \kappa$, with at least one level of exponentiation as described in section \ref{Sec-modular-exponentiation-over-Finite-Fields}, are required. It is also assumed that the key generator ensures that a prospective owner of the pertinent keys is provided with an abundance of options for generating multivariate one-to-one mappings, whose inverse mappings are known only to the owner.  
   
 In public key cryptography, the size of the set {\small{$\left \{ F(\xx,\, \boldomega) : \,\boldomega \in \scalars^{\kappa} \right\}$}} must be large, such as perhaps exponential in $\nu^{c}$, for some fixed $c > 0$, for each $ \xx \in \scalars^{\mu}$, while maintaining $F(\xx,\, \boldomega)$ as a secret to the public,
 for IND-CCA and IND-CPA security, whichever is relevant.  Under the assumption of no mistrust, the padding message $\boldomega$ can be negotiated for mutual agreement by sender and receiver, for ascertaining data integrity, in public key cryptography.

For digital signature, the main security issue is the anonymity of the secret keys
{\small{$F(\xx,\, \boldomega)$
$ = $
$\bglb\basel{f}{1}(\xx,\, \boldomega), \, \ldots,\,\basel{f}{L}(\xx,\, \boldomega)\bgrb$},} 
which are registered with a trusted authentication verifier (TAV). The size of the set $\{\boldomega \in \mygrp^{\kappa} \, :\, H(\zz,\, \xx,\, \boldomega) = \bolddelta\}$, where $\zz \in \mygrp^{\lambda}$, $\xx \in \mygrp^{\mu}$ and $\bolddelta \in \eltSet^{\tau}$, must be large, such as perhaps exponential in $\nu^{c}$, for some fixed $c > 0$, whenever the set in the discussion is nonempty, and the ratio of the number of elements in  $\{(\boldomega,\, \zz) \in \mygrp^{\kappa+\lambda} \, :\, H(\zz,\, \xx,\, \boldomega) = \bolddelta\}$ to that in $\{(\boldomega,\, \zz) \in \mygrp^{\kappa+\lambda} \, :\, H(\zz,\, \xx,\, \boldomega) = \bolddelta\}$ 
 $\cap$
 $\{(\boldomega,\, \zz) \in \mygrp^{\kappa+\lambda} \, :\, R(\zz) = F(\xx,\, \boldomega)\}$ must be large,  if $H(\zz,\, \xx,\, \boldomega) $ occur in the public key signature verification table $\SignatureVerificationTable$, whenever the stated sets are nonempty, such as perhaps exponential in $\nu^{c}$, for some fixed $c > 0$ and any fixed $\bolddelta \in \eltSet^{\tau}$, admissible plain message $\xx \in \mygrp^{\mu}$ and admissible padding message $\boldomega \in \mygrp^{\kappa}$. The admissibility of the padding message $\boldomega \in \mygrp^{\kappa}$ is that $R(\boldomega) = \boldomega'$, where $\boldomega' \in \mygrp^{K}$ is the extra padding message agreed upon by the signer with TAV, for the particular reserved transaction, as a first step in the process of generating the signature. The hidden or secret keys $F(\xx,\,\boldomega)$ must not be made known to the public, but, in the digital signature scheme, are shared with the TAV, besides the signer, in order to ensure existential unforgeability. For claiming the authenticity of a signature, the claimant needs to produce $\boldepsilon$, $\bolddelta$, $\xx$, $\zz$ and $\boldomega$, for passing the tests of TAV, without knowing the information in the signature authentication verification table $\SignatureAuthenticationTable$, which is registered with TAV. The signer must assert with TAV,
 by means of $\boldomega'$ and other protocol agreements, regarding authorization of a signature.
 Thus, the signer is protected by the prudence and unbiasedness of TAV.
   

\section{Conclusion and Summary}

In this paper, a new public key data encryption method is proposed, where the plain and encrypted messages are arrays. The method can also be used for digital certificate or digital signature applications. For security protocols of the application layer level in the OSI model, the methods described in this paper are useful. In the regular protocols like TLS and IPSec [\cite{StallingsCNS:2011} and  \cite{StallingsNSE:2011}], the traditional methods, requiring only small space for the keys and algorithms, are employed. The key generation algorithm is particularly simple, easy and fast, facilitating changes of keys as frequently as required, and fast algorithms for polynomial multiplication and modular arithmetic [\cite{BZ:2010} and  \cite{Pan:1992}], whenever appropriate, can be adapted in the encryption and decryption algorithms.

\end{document}